\documentclass[aps,pre,reprint, amsmath, amssymb,superscriptaddress]{revtex4-1}
\usepackage{morefloats}
\usepackage{soul}
\usepackage{bm}
\newcommand{\beq}{\begin{equation}}
\newcommand{\eeq}{\end{equation}}

\usepackage[retainorgcmds]{IEEEtrantools}
\usepackage{graphicx,tikz,placeins}
\usepackage{mathrsfs}
\usepackage{wasysym}
\usepackage{amsmath,amssymb,amsfonts,physics}
\usepackage{color}
\usepackage{float}
\usepackage{lipsum}
\usepackage{comment}
\usepackage[normalem]{ulem}

\usepackage{times,txfonts}
\usepackage{nicefrac}
\usepackage[colorlinks=true,linkcolor=blue,urlcolor=blue,citecolor=blue,pdfusetitle]{hyperref}
\usepackage{physics}
\usepackage{soul}
\newcommand{\mom}[1]{\langle#1\rangle}

\begin{document}

\title{Steady-state heat engines driven by finite reservoirs}

\author{Iago N. Mamede}
\affiliation{Universidade de São Paulo,
Instituto de Física,
Rua do Matão, 1371, 05508-090
São Paulo, SP, Brazil}

\author{Saulo V. Moreira}
\affiliation{School of Physics, Trinity College Dublin, College Green, Dublin 2, D02 K8N4, Ireland}

\author{Mark T. Mitchison}
\affiliation{School of Physics, Trinity College Dublin, College Green, Dublin 2, D02 K8N4, Ireland}
\affiliation{Department of Physics, King’s College London, Strand, London, WC2R 2LS, United Kingdom}

\author{Carlos E. Fiore}
\affiliation{Universidade de São Paulo,
Instituto de Física,
Rua do Matão, 1371, 05508-090
São Paulo, SP, Brazil}


\begin{abstract}

We provide a  consistent thermodynamic analysis
of stochastic thermal engines driven by finite-size reservoirs, which are in turn coupled to infinite-size reservoirs. We consider a cyclic operation mode, where the working medium couples sequentially to hot and cold reservoirs, and a continuous mode with both reservoirs coupled simultaneously.
We derive an effective temperature for the finite-size reservoirs determining the  entropy production for two-state engines in the sequential coupling scenario, and show that finite-size reservoirs can meaningfully affect the power when compared to infinite-size reservoirs in both sequential and simultaneous coupling scenarios.
We also investigate a three-state engine comprising two interacting units and optimize its performance in the presence of a finite reservoir. 
Notably, we show that the efficiency at maximum power can exceed the Curzon-Ahlborn bound with finite reservoirs.
Our work introduces tools to optimize the performance of nanoscale engines under realistic conditions of finite reservoir heat capacity and imperfect thermal isolation.

\end{abstract}

\maketitle

\section{Introduction}
The understanding of non-equilibrium phenomena in nano-scale systems has significantly advanced since the advent of stochastic thermodynamics, which generalized thermodynamics to the context of few particles and stochastic trajectories~\cite{sekimoto_first,Seifert_2012}  and  has found applications in the realm of information processing~\cite{parrondo}, chemical reaction networks \cite{PhysRevX.6.041064},
active matter \cite{PhysRevX.7.021007,PhysRevX.9.021009} and numerous experimental set-ups \cite{RevModPhys.85.1421,ciliberto2017experiments}. 
Notably, it has provided new insights and tools to tackle {\it heat engines} at the microscopic level~\cite{Hondou2000, Schmiedl2008, Zhang2010, Holubec2014, Chiang2017, Argun2017, vonLindenfels2019,VANDENBROECK20156}. A heat engine, enabling work extraction by placing a system in contact with hot and cold reservoirs, has been a fundamental concept since the Industrial Revolution, with Sadi Carnot's vision of cyclic operation providing the basic description of macroscopic energy conversion ever since~\cite{carnot1978reflexions, callen}.
Recent studies have investigated some of the novel issues arising for engines operating at the nanoscale, such as optimization under realistic constraints, e.g.~finite times
\cite{Tu2008,ciliberto2017experiments,blickle2012realization}, the effect of different optimization strategies (power or efficiency)~\cite{Dechant2017, PhysRevE.96.052135,felipe}, and the trade-off between performance, dissipation and fluctuations \cite{karel2016prl, PhysRevLett.117.190601, Pietzonka2018,Miller2021}.
In general, most  investigations take one of three approaches for heat engines:
 periodically driven working media \cite{Brandner2015, Bauer2016, Brandner2020, CLEUREN2020122789,mamede2022}, sequential coupling to hot and cold thermal baths \cite{Schmiedl2008, PhysRevLett.105.150603,barato,PhysRevE.96.052135,PhysRevE.81.041106,Iyyappan_2019,felipe,filho21}, or simultaneous contact with both baths \cite{PhysRevLett.102.130602,gatien,mamede2023thermodynamics}, the latter being commonly  viewed as the limit of sequential operation under rapid switching \cite{PhysRevE.96.052135,busiello2021dissipation,felipe}.
 
The vast majority of recent works have considered infinite-size reservoirs, which is a good approximation when the reservoir is large. 
However, when at least one of the reservoirs has a finite heat capacity, its temperature changes whenever it exchanges heat with the system. Recent works have evidenced thermodynamic implications of these temperature fluctuations~\cite{PRXQuantum.2.030202, PhysRevE.104.L022103, PhysRevLett.131.220405,uzdin2018markovian,PRXQuantum.2.010340} and their effects on transport properties~\cite{Schaller2014, matern2024thermoelectriccoolingfinitereservoir}. Heat engines coupled to finite thermal reservoirs have also been studied under the assumption of a fixed total energy~\cite{Izumida2014, PhysRevE.90.062140, PhysRevE.94.012123,  Ma2020, Yuan2022}. 
This limits the engine's operation time because the reservoirs eventually equilibrate with each other, as has been observed in isolated systems such as ultracold atomic gases~\cite{Brantut2013}. However, in many situations the finite reservoir is not fully isolated from its environment, e.g.~as in recent experiments where finite electron reservoirs are observed to equilibrate with their phononic environment on a timescale much longer than that of their own fast, internal thermalisation~\cite{karimi2020reaching, Champain2024}. This coupling to an external environment opens the interesting possibility of steady-state engine operation driven by finite-size reservoirs, but a detailed analysis of this scenario is still lacking.

In this paper, we fill this gap by proposing stochastic descriptions
of heat engines
in contact with finite thermal reservoirs, which are in turn coupled to external thermostats such that a steady state is reached.
In Scenario I, we introduce a cyclic heat engine  composed of two isothermal steps, each one operating at a finite time duration $\tau_1$ and $\tau_2$, plus two (instantaneous) adiabatic
steps. The total duration of the cycle is given by $\tau = \tau_1 + \tau_2 + \tau_3$, where $\tau_3$ is an additional time duration, in which the system is disconnected from both reservoirs, accounting for the time it takes to restart the cycle.
 At least one out of the two reservoirs will be treated as finite.
{We assume that $\tau_3$, in which the finite-size reservoir has its initial energy replenished through its contact with its infinite-size reservoir, be 
long enough such that $\tau \gg \tau_1+\tau_2$}.
 The thermalization of the finite-size reservoir, in turn, is considered to happen in a timescale much smaller than $\tau_1$, such that its temperature fluctuates due to the exchanged heat with the system.
On the other hand, when the period is very short, $\tau\ll 1$, or in the case of a simultaneous contact between the system and both thermal baths, the assumption of finite-size reservoir being replenished  will not be fulfilled. For this reason, we propose a complementary description in Scenario II, where we relax the assumption of {slower thermalization between the finite-size and its infinite reservoir
and we assume instantaneous  thermalization  as the finite-size reservoir exchanges heat with both system and infinite reservoir}.
We derive thermodynamic quantities for such heat engines coupled to finite-size reservoirs in both scenarios, by considering two- and three-state systems, which imply significant differences in engine performance 
when compared to the infinite-reservoir case.

\section{Thermodynamics of finite reservoirs}
{\it Scenario I.}---We consider a generic $N$-state system
with energies $\{\varepsilon_{ i}\}_{i=0}^{N-1}$ and sequential contact with the $\nu$-th thermal reservoir. The system is placed in contact with the cold (hot) finite reservoir during the time interval $0 < t \le \tau_1$ (stage 1), and then with the hot (cold) infinite reservoir when $\tau_1< t\leq\tau_1+ \tau_2$ (stage 2).
The switching between thermal baths occurs at $t=\tau_1$, and  a periodic cycle is completed at $t=\tau \gg \tau_1 + \tau_2$.
The process is then restarted, with
the finite reservoir energy restored to its initial value. The rate of transitions from state $j$ to $i$ is expressed via the Kramers relation, $\omega_{ij}=\Gamma \exp\{-\beta_{ij}\Delta\varepsilon_{ij}/2\}$, where $\Delta\varepsilon^{}_{ij}$ is the energy difference between states $i$ and $j$, $\Gamma$ is the coupling strength between the system and reservoir, and $\beta_{ij}$ denotes the reservoir's inverse temperature before the transition takes place.

When the system is coupled to an infinite bath at temperature $T$, as is the case in stage 2, we have
$\beta_{ij}=\beta_{}=T_{}^{-1}$ (we set $k_B=1$). However, when the system is coupled to a finite-size reservoir, the temperature of the latter fluctuates, as it exchanges discrete amounts of heat with the system. Thus, assuming that the finite reservoir is characterized by
 a constant heat capacity $C>0$, and given that the total initial energy of the finite reservoir and system is fixed and given by $\mathcal{E}$, the reservoir's temperature $\beta^{}_{ij}$ associated with the transition rate from state $j$ to $i$ will differ from $\beta^{}_{ji}$ for the reverse transition from $i$ to $j$.
 The local detailed balance conditions imply that the transition rates satisfy
\begin{equation}
    \centering
    \log\frac{\omega_{ij}^{}}{\omega_{ji}^{}}={-\frac{1}{2}(\beta^{}_{ij}+\beta^{}_{ji})\Delta\varepsilon^{}_{ij}}.
    \label{LDBFinite}
\end{equation}

At each isothermal subprocess, when the system is in contact with the $\nu$-th reservoir, the system dynamics is given by the following master equation,
\begin{equation}
    \centering
    \dot{p}^{}_i(t)= \sum_{j\neq i} \omega^{(\nu)}_{ij}p^{}_{j}(t)-\omega^{(\nu)}_{ji}p^{}_{i}(t) =\sum_{j\neq i}J^{(\nu)}_{ij},
    \label{meMain}
\end{equation}
  where $p_{i}^{}(t)$ denotes the probability that the system is in state $i$ at time $t$, and we use a superscript $\nu$ to indicate quantities specific to reservoir $\nu$. We denote by $\vb{p}(t) \equiv (p^{}_0(t), p^{}_1(t), \dots, p^{}_{N-1}(t))^{\rm T}$ the vector containing the probabilities $p_i$ which is
  consistent with its continuity along
   switchings.


The amount of power in each stage is given by
${\mom{\mathcal{P}(t)}}=\sum_i(\varepsilon^{}_{\nu' i}-\varepsilon^{}_{\nu i})\delta(t-\tau_\nu)p_i(t)$
\cite{Crooks1998,Filho_2024} 
where $\{\varepsilon_{\nu i}\}_{i=0}^{N-1}$ are the system energies when it is coupled to the $\nu$-th reservoir, with $\nu(\nu')=1(2)$ and $2(1)$. The integration  over a complete cycle, together with the above boundary conditions, gives
\begin{equation}
    {\mom{\cal P}}=-\frac{1}{\tau}\sum_{i}(\varepsilon^{}_{2i}-\varepsilon^{}_{1i})\left(p_i(\tau_1)-p_i(0)\right).
\end{equation}
Likewise, the rate of average heat exchanged with the $\nu$-th bath reads $\langle {\dot Q}_\nu\rangle= \sum_{i<j}(\varepsilon^{}_{\nu i}-\varepsilon^{}_{\nu j})\overline{J}^{(\nu)}_{ij}/\tau$, where $\overline{J}^{(\nu)}_{ij}$ denotes  the cycle-averaged flux, the expressions for which are given in the Supplemental Information (SI).

{\it Scenario II.}---The $N$-state system is  considered to be simultaneously coupled to two finite-size reservoirs as $\tau \ll 1$. 
In turn, each finite-size reservoir, which is now described by  the time-dependent temperature ${\tilde T}_\nu (t)$, is in contact with an infinite-size reservoir at temperature $T_\nu$.
The time evolution of system is given by
\begin{equation}
    \centering
    \dot{p}_i(t)=\sum_{\nu}\sum_{j\neq i}[\omega_{ij}^{(\nu)}(t)p_j(t)-\omega_{ji}^{(\nu)}(t)p_i(t)]=\sum_{\nu}\sum_{j\neq i}J_{ij}^{(\nu)}(t),
    \label{ME}
\end{equation}
where $\omega_{ij}^{(\nu)}(t) = \Gamma \exp(-\Delta \varepsilon^{(\nu)}_{ij}/2\tilde T_\nu(t))$ is the transition rate from state $j$ to $i$.
Given the heat capacity $C_\nu$ of the $\nu$-th finite-size reservoir, which we assume to be constant,
the time evolution of ${\tilde T}_\nu (t)$ is given by
\begin{equation}
    \centering
    C_\nu\dv{\tilde{T}_\nu(t)}{t}=-\mom{\dot{Q}_\nu(t)}-\kappa_\nu(\tilde{T}_\nu(t)-T_\nu),
    \label{ODETemp}
\end{equation}
where $\mom{\dot{Q}_\nu(t)}=\sum_{i<j}(\varepsilon^{}_{\nu i}-\varepsilon^{}_{\nu j})J^{(\nu)}_{ij}(t)$ is the average heat exchanged between the $\nu$-th reservoir and the system and $\kappa_\nu$ is the thermal conductivity between finite and infinite reservoirs, which is also assumed to be constant.
In the long time limit, the solution to Eq.~\eqref{ODETemp} reads
$\tilde{T}^*_\nu=T_\nu-\mom{\dot{Q}_\nu}/\kappa_\nu$, implying that $\tilde{T}^*_2<T_2$ and $\tilde{T}^*_1>T_1$ in the heat engine regime {and the other way around
for the heat pump regime}. Furthermore, from the first law, the power is given by
 \begin{equation}
     \centering
     \mom{\mathcal{P}}=-\mom{\dot{Q}_1}-\mom{\dot{Q}_2}=\kappa_1(\tilde{T}^*_1-T_1)+\kappa_2(\tilde{T}^*_2-T_2).
 \end{equation}

\section{Main Results}
{\it Two-state engines.}---We now consider the simplest engine design, composed of a two-state unit with energies $\{0,\varepsilon _\nu\}$, corresponding to the occupation of the lower and upper state, respectively. Different operation regimes stem from the fact that the upper state assumes different values,   $\varepsilon_1$ and $\varepsilon_2$,  according to whether it is  in contact with reservoir 1 or 2, respectively.
Specifically, for infinite-size reservoirs, previous studies \cite{PhysRevE.96.052135,felipe} predict a heat-engine for $\varepsilon_2-\varepsilon_1>0$ or heat pump for $\varepsilon_2-\varepsilon_1<0$, whose power increases as the difference of temperatures between thermal baths grows.

\begin{figure*}
    \centering
    \includegraphics[scale=0.26]{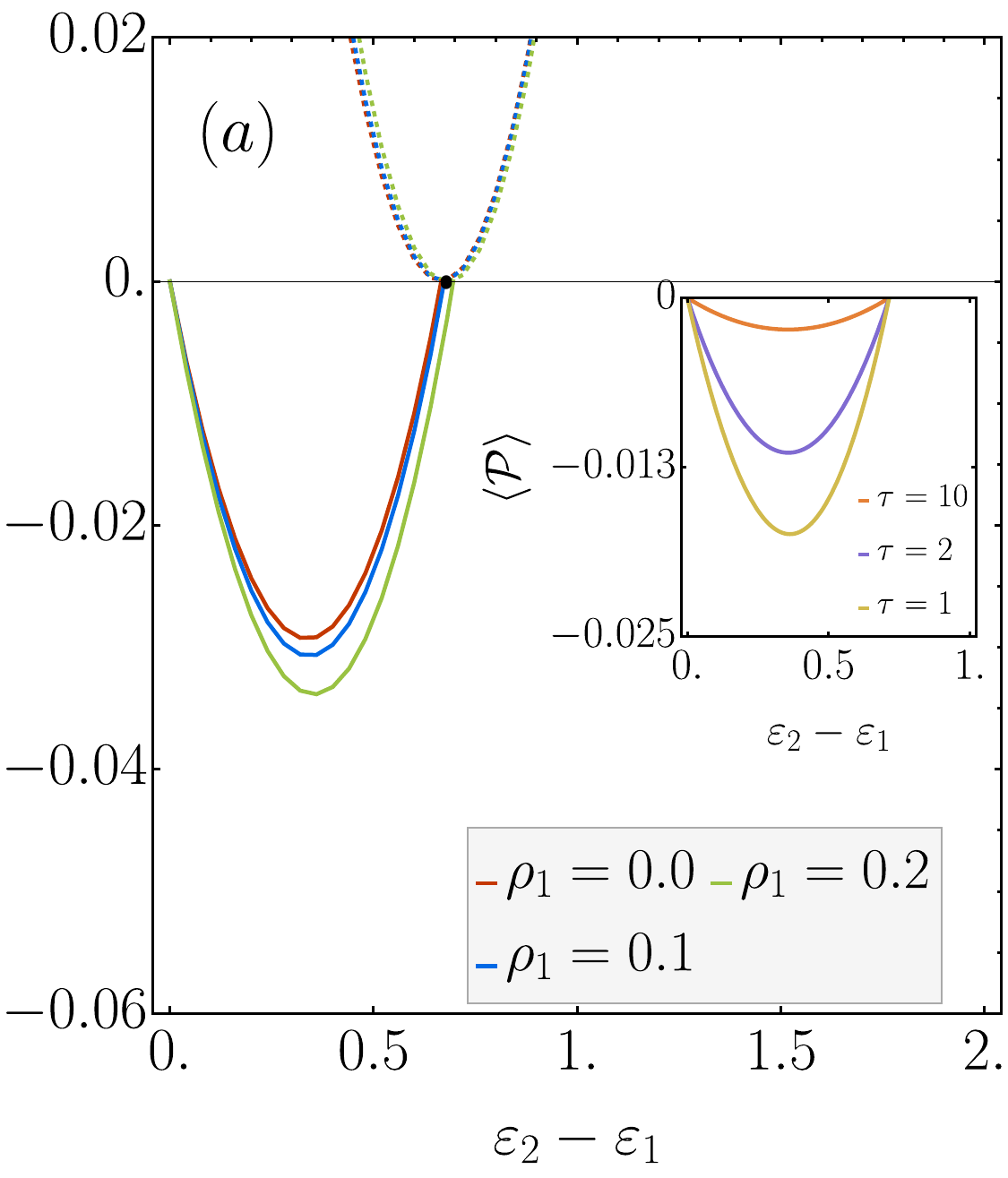}
    \includegraphics[scale=0.26]{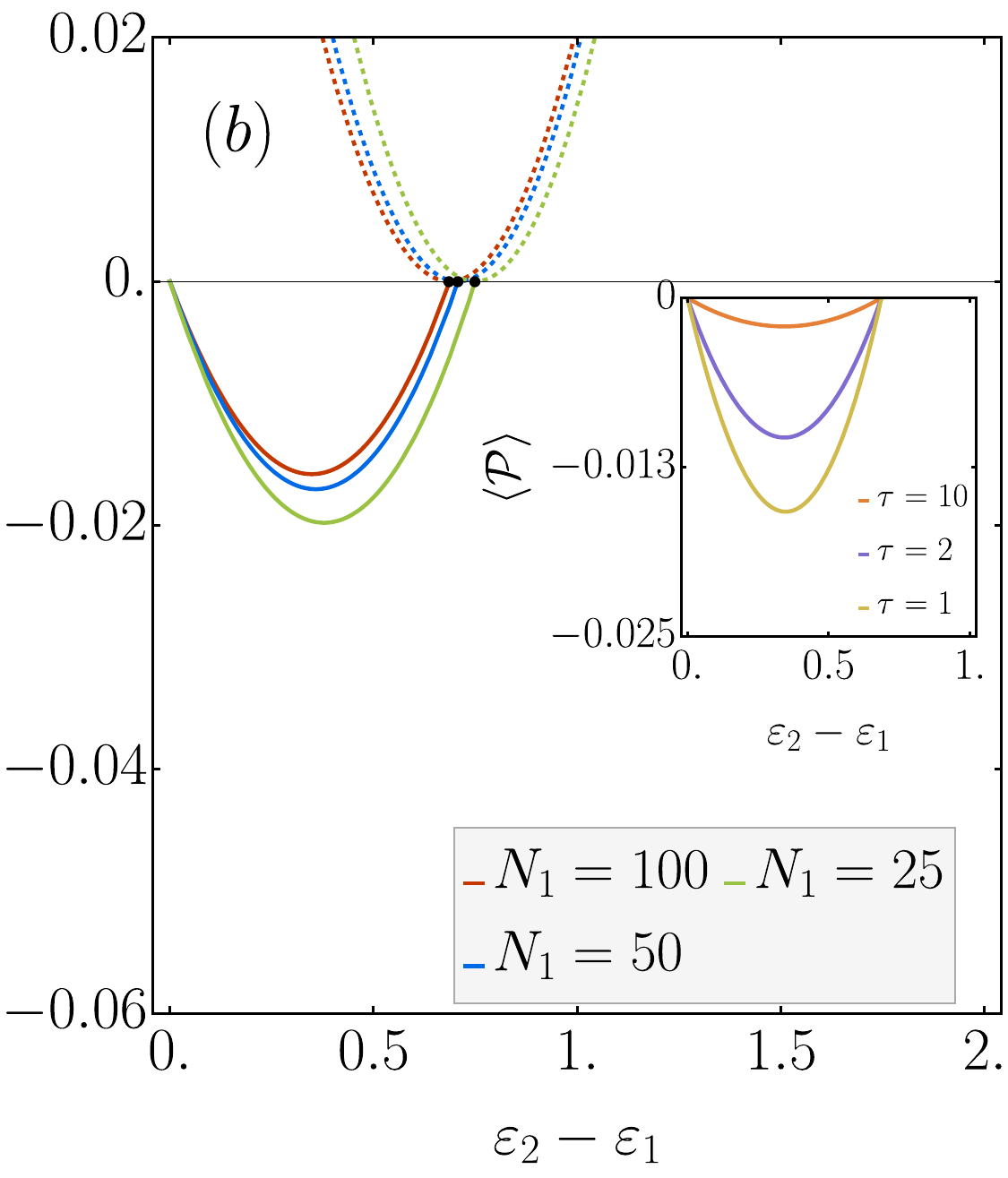}
    \includegraphics[scale=0.26]{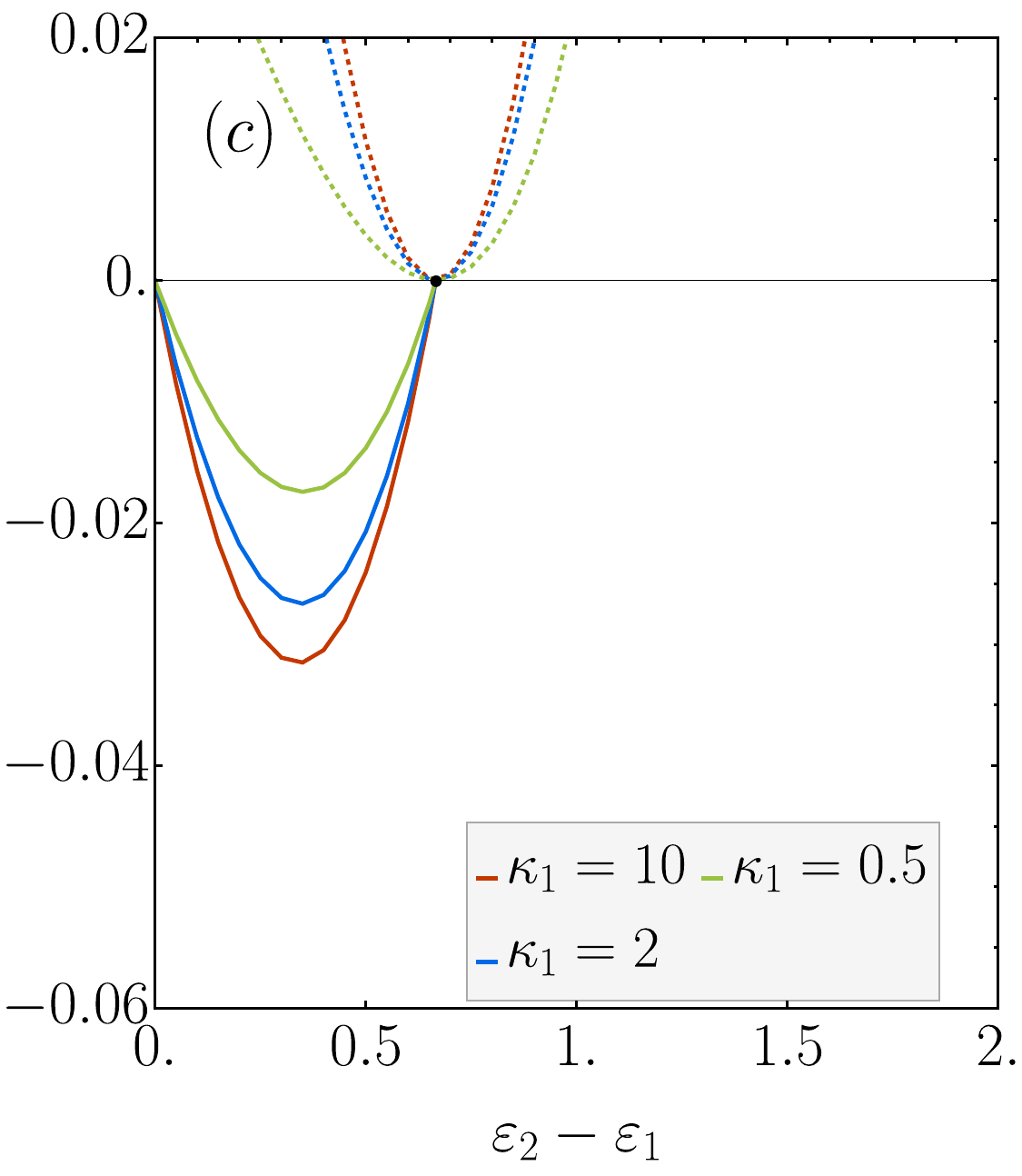}
    \caption{Average power $\mom{\mathcal{P}}$ (solid lines) and entropy production rate $\mom{\dot{\sigma}}$ (dashed lines) as a function of $\varepsilon_2 - \varepsilon_1$ in Scenario I, for a two-state engine placed in contact with a cold finite-size reservoir and a hot infinite-size reservoir. We consider that the finite-size reservoir is characterized by (a) a constant heat capacity and (b) an ensemble of two level systems. 
    Insets: Power versus $\varepsilon_2-\varepsilon_1$ for different values of $\tau$. In (c), we plot $\mom{\mathcal{P}}$ (solid lines) and $\mom{\dot{\sigma}}$ (dashed lines) as a function of $\varepsilon_2 - \varepsilon_1$ in Scenario II.
}
    \label{fig_General}
\end{figure*}

We first characterize the finite-size reservoirs in both scenarios through a finite and constant heat capacity, $C_\nu$. From now on, for the sake of generality, we will consider that, in Scenario I, either the cold (for $\nu=1$) or the hot reservoir (for $\nu=2$) can be finite. Therefore, given the total energy $\mathcal{E}_\nu$ of the finite-size reservoir and system, we have that
\begin{equation}\label{finitedef_1}
\beta_{10}^{(\nu)}=\beta_\nu,\quad \beta_{01}^{(\nu)}=\beta_\nu\left(1-\rho_\nu\right)^{-1},
\end{equation}
 if the system is empty at the beginning of the stage, where $\beta_{\nu}\equiv C_{\nu}/\mathcal{E}_\nu$ and $\rho_\nu\equiv\varepsilon_\nu/\mathcal{E}_\nu$. Likewise,
\begin{equation}\label{finite_def_2}
\beta_{10}^{(\nu)}=\beta_\nu\left(1+\rho_\nu\right)^{-1}, \quad
 \beta_{01}^{(\nu)}=\beta_\nu,
 \end{equation}
 if the system is occupied at the beginning of the stage. 
 For example, if the cold reservoir is finite, the occupation of the system at $t=\tau_1+\tau_2$, which is a stochastic variable, will determine if its temperature will fluctuate between the values $\beta_1$ and $\beta_1(1-\rho_1)^{-1}$ from Eq.~\eqref{finitedef_1}, or between $\beta_1(1+\rho_1)^{-1}$ and $\beta_1$ from Eq.~\eqref{finite_def_2}.
Thus, the concept of \textit{entropic temperature} from Ref.~\cite{PhysRevLett.131.220405}, defined as a (stochastic) temperature dictating the entropy production of a two-level system coupled to a finite-size reservoir along a single trajectory, is useful here. Since the total energy of the system and reservoir is fixed at the beginning of each stage, the inverse entropic temperature reads
 \begin{equation}
     \beta_e^{(\nu)} = \frac{1}{\epsilon} \log \frac{\omega_{10}^{(\nu)}}{\omega_{01}^{(\nu)}} = \frac{\beta_{10}^{(\nu)} + \beta_{01}^{(\nu)}}{2}.
 \end{equation}
 Indeed, we will see that the average entropy production rate in Scenario I can be written in terms of the entropic temperatures of each reservoir. We define the average entropic (effective) temperature by
\begin{equation}\label{effbeta}
     \beta_{\text{eff}}^{(\nu)} \equiv p_{0\nu}^{} \beta_{\text{e}0}^{(\nu)} + p_{1\nu}^{} \beta_{\text{e}1}^{(\nu)},
 \end{equation}
 where $p_{0\nu}^{}$, $p_{1\nu}^{}$ denote the probability that the system is empty, with the reservoir's associated entropic temperature $\beta_{\text{e}0}^{(\nu)}$, or occupied, with the reservoir's corresponding entropic temperature $\beta_{\text{e}1}^{(\nu)}$, when it comes into contact with the $\nu$-th reservoir, respectively.
Specifically, the entropy production $\mom{\dot{\sigma}}$ at the non-equilibrium steady-state in each scenario assumes the form $\mom{\dot{\sigma}}=-\sum_\nu\xi_\nu\mom{\dot{Q}_\nu}$, with
\begin{align}
    \centering
    \xi_\nu=\begin{cases}
\displaystyle \beta^{(\nu)}_{{\rm eff}}~~~~~~~~~~~~~\text{(Scenario I)}\\
        \\
        \displaystyle\frac{1}{\tilde{T}^*_\nu}~~~~~~~~~~~~~\text{(Scenario II)}.
    \end{cases}
\end{align}
The average heat rate can be expressed as
$\langle {\dot Q}_\nu\rangle=\varepsilon_\nu\mathcal{J}_s$, where  $\mathcal{J}_s=\overline{J}^{(1)}_{10}$ in Scenario I,
and $\mathcal{J}_s=J^{(1)}_{10}$ in Scenario II.
The efficiency $\eta$, given by the ratio between  
$\langle {\cal P}\rangle=(\varepsilon_1-\varepsilon_2)\mathcal{J}_s$
and $\mom{\dot{Q}_2}$, acquires the simple form $\eta=1-\varepsilon_1/\varepsilon_2$, implying  that the ideal
regime of operation is reached when $\varepsilon_1/\varepsilon_2=\xi_2/\xi_1$.

    \label{finite_def_1}
Going beyond this description, {we also consider an alternative approach for Scenario I where the heat capacity of the finite reservoir is energy-dependent. Here,} the finite-size reservoir is constituted of
$N_\nu$ non-interacting units (akin to the system), each of them also occupying two states with individual energies $0$ or $\varepsilon_\nu$. 
The latter approach has a long tradition in equilibrium statistics \cite{callen,salinas2001introduction}, in which the (microcanonical)
temperature is evaluated via the density of states in which  $N_\nu$ and $\mathcal{E}_{\nu}$ held fixed. When the system is empty when it comes into contact with the reservoir, we have
\begin{equation}
\beta_{10}^{(\nu)}=\varepsilon_\nu^{-1}\mathcal{F}\left(\frac{N_\nu\varepsilon_\nu}{\mathcal{E}_{\nu}}\right), \quad
\beta_{01}^{(\nu)}=\varepsilon_\nu^{-1}\mathcal{F}\left(\frac{(N_\nu-1)\varepsilon_\nu}{\mathcal{E}_{\nu}}\right),
\label{Temp_TLS}
\end{equation}
where $\mathcal{F}(x)\equiv\log[(1-x)/x]$.
In each approach, fluctuations of $\mathcal{E}_v$ or $N_\nu$ lead the reciprocal temperature to vary between $ \beta_{10}^{(\nu)}$ and $ \beta_{01}^{(\nu)}>\beta_{10}^{(\nu)}$. Therefore, $\beta_{01}^{(\nu)}\rightarrow \beta_{10}^{(\nu)}$ when $C_\nu\to\infty$, $\mathcal{E}_{\nu}\to\infty$ ($\rho_\nu\rightarrow 0$),  or $N_\nu \rightarrow \infty$.



We obtain exact expressions for thermodynamic quantities such as power, heat, dissipation and efficiency in the Supplemental Information (SI). 
These expressions are significantly different from the situation where all reservoirs are infinite~\cite{felipe,PhysRevE.96.052135}, the latter case being recovered for $\rho_\nu\rightarrow 0,N_\nu\rightarrow \infty$ (Scenario I) and $\kappa_\nu\rightarrow \infty$ (Scenario II). 
{Fig.~\ref{fig_General} depicts} these results by considering the cold reservoir finite ($\rho_1 \neq 0$) and the hot reservoir infinite ($\rho_2 = 0$).
We plot the power and entropy production as a function of $\Delta \varepsilon = \varepsilon_2 -\varepsilon_1$, for  $\tau_1=\tau_2$ and different reservoirs sizes, as quantified by $\rho_1$ in panel (a) and $N_\nu$ in panel  (b). 
First, we note that the engine regime is always delimited by $\Delta\varepsilon=0$ and $(\Delta\varepsilon)_{\sigma_{\text{min}}}$, i.e. the value of $\Delta\varepsilon$ corresponding to the entropy production minimum \cite{forao2025}. We see that $(\Delta\varepsilon)_{\sigma_{\text{min}}}$ is affected by the finiteness of the reservoir, meaning different values of $\rho_1$ and $N_1$ in Figs.~\ref{fig_General}(a) and~\ref{fig_General}(b), respectively.
For small $\rho_1$ and $\rho_2$ and defining  $\epsilon = \varepsilon_1 + \varepsilon_2$, it is approximately given by
\begin{equation}\label{Deltaepsilon}
    (\Delta\varepsilon)_{\sigma_{\text{min}}} \approx \epsilon \frac{\beta_1 - \beta_2}{\beta_2+\beta_1} + \epsilon\beta _1 \beta _2 \frac{\rho _1-\rho _2}{(\beta_1+\beta_2)^2}.
\end{equation}
We note that the first term on the right-hand side of Eq.~\eqref{Deltaepsilon} is the value of $(\Delta\varepsilon)_{\sigma_{\text{min}}}$ when both reservoirs are infinite ($\rho_1 = \rho_2=0$).
Thus, the engine regime is enlarged when the cold reservoir is finite ($\rho_1 \neq 0$) and the hot is infinite ($\rho_2 = 0$), as shown in Fig.~\ref{fig_General}(a).
Conversely, the engine regime is shortened when the cold reservoir is infinite ($\rho_1 = 0$) and the hot is finite ($\rho_2 \neq 0$).
Furthermore, by comparing Figs.~\ref{fig_General}(a)/(b) and~\ref{fig_General}(c), we see different behaviours of the power in relation to the finiteness of the reservoir in Scenarios I and II. Specifically, the power increases as $\rho_1$ ($N_1$) increases (decreases) in Fig.~\ref{fig_General}(a) (Fig.~\ref{fig_General}(b)),
while it decreases as $\kappa_1$ decreases in Fig.~\ref{fig_General}(c). As a result, in both descriptions for the finite-size reservoir, as captured by $\rho_1$ and $N_1$, increasing the finiteness of the cold reservoir leads to an increase in power in Scenario I.
However, when the cold reservoir is infinite ($\rho_1=0$) and the hot is finite ($\rho_2\neq 0$), the power decreases as the reservoir's finiteness increases for both descriptions in Scenario I.

Such difference in the way the power behaves, depending on which reservoir, cold or hot, is finite, can be understood by noting that the system presents a single independent flux and that the fluctuation of the inverse temperature between the values in Eq.~\eqref{finitedef_1} or in Eq.~\eqref{finite_def_2}, depending on the state of the system when it comes into contact with the reservoir (empty or occupied), leads to different entropic temperatures of the finite-size reservoir in each cycle, and to the definition of the inverse effective temperature in Eq.~\eqref{effbeta}.
Specifically, when the finite-size reservoir is cold, the inverse effective temperature $\beta_{\text{eff}}^{(1)}$
is \textit{colder} than $\beta_1$.
In turn, when both reservoirs are infinite, the inverse temperature of the cold reservoir becomes fixed and equal to $\beta_1$. Thus, when $\beta_{\text{eff}}^{(1)}$ is larger than $\beta_1$,
we see an increase in power extraction on average. In the same vein, when the hot reservoir is finite, $\beta_{\text{eff}}^{(2)}$ is larger (and therefore colder) than $\beta_2$.
As a result, we see a reduction of the extracted power, on average (see SI).  {Although derived within the general context of two-state systems, the above findings are also extended beyond the two-state case, strongly suggesting their general occurrence.
 Finally, in relation to Scenario II, the difference in temperature of the finite-size reservoirs, $\tilde{T}_2^*-\tilde{T}_1^*$, always decreases in comparison with the difference in temperature when both reservoirs are infinite, $T_2 - T_1$. Therefore, the heat engine with finite-size reservoirs in the simultaneous coupling scenario always presents an inferior performance in terms of power extraction, irrespective of the system details, provided $\langle{\cal P}\rangle$ increases with the difference of temperatures.}

{\it Beyond  two-state engines.}---In the previous example, we saw that finite-size reservoirs can significantly affect the average power. However, 
the efficiency $\eta$ is not affected due to the fact that  two-state engines present only one independent flux.
Thus, to investigate the effect of finite-size reservoirs on the efficiency,
we consider a more complex heat engine model, which includes the interaction between different units, and explore different optimization routes.
A similar model consisting of many interacting nanomachines has been shown to exhibit collective behavior and a phase transition for different topology of interactions~\cite{gatien, mamede2023thermodynamics}.


Specifically, we consider two interacting units \cite{felipe}, where each unit 
can also be in a lower  or upper  state, with individual energies $0$ and $\varepsilon_\nu$, respectively, with an interaction term $V_\nu$ added if two units are in different states.
By characterizing microscopic states in terms of total particle number $i$ occupying the upper state, the dynamics
reduces to a three state system  (labeled by $i=0,1$ or $2$)
with associate 
energies $0,V_\nu+\varepsilon_\nu$ and $2\varepsilon_\nu$, respectively \cite{gatien,mamede2023thermodynamics}. 
Different aspects,
such as behaviour of power, efficiency, the existence of different regimes of operation and optimization routes, depend on the rich interplay between temperatures, individual and interaction energies. In particular, for two infinite-size reservoirs, the crossover between heat engine and heat pump regimes  is featured with maximum efficiencies $\eta_{ME}$, which are equal to Carnot efficiency
provided $\varepsilon_1/\varepsilon_2=V_1/V_2=T_1/T_2=r^*$, 
otherwise maximum efficiencies satisfy ${\eta}_{ME}<\eta_c$. 

We now consider a heat engine operating as described for Scenario I, but with the three-level system.
Thus, following the same logic as before, if the system is in the state $i=0$ when it is put in contact with the $\nu$-th reservoir with energy $\mathcal{E}_\nu$, the inverse temperatures $\beta_{ij}^{(\nu)}$ will fluctuate as discrete amounts of energy are exchanged between system and reservoir:
\begin{align}
    \centering
    \beta^{(\nu)}_{10}&=\beta_\nu,\nonumber\\ \beta^{(\nu)}_{01}&=\beta^{(\nu)}_{21}=\beta_\nu\left[1-({\rho_\nu+\mathcal{V}_\nu})\right]^{-1},\\ 
    \beta^{(\nu)}_{12}&=\beta_\nu\left(1-{2\rho_\nu}\right)^{-1},\nonumber
\end{align}
where $\rho_\nu\equiv\varepsilon_\nu/\mathcal{E}_\nu$ and $\mathcal{V}_\nu\equiv V_\nu/\mathcal{E}_\nu$. The case where both reservoirs are infinite corresponds to $\rho_\nu=\mathcal{V}_\nu=0$ \cite{felipe}. The system dynamics is described in detail in the SI. Furthermore, we obtain exact expressions
for the
temperatures ${\tilde T}_\nu$
in the SI for Scenario II. To sum up, in both scenarios,  $\mom{{\cal P}}$
and $\langle {\dot Q}_2\rangle$ can be expressed as
\begin{equation}\label{A_Label}
  \begin{split}
    \mom{\mathcal{P}}&=-(\varepsilon_1-\varepsilon_2)\mathcal{K}_1+(V_1-V_2)\mathcal{K}_2,\\
   \mom{\dot{Q}_2}&=-(V_2+\varepsilon_2)\mathcal{J}_{1}+(V_2-\varepsilon_2)\mathcal{J}_{2},
  \end{split}
\end{equation}
where $\mathcal{K}_1\equiv\mathcal{J}_2+\mathcal{J}_1$ and $\mathcal{K}_2\equiv\mathcal{J}_2-\mathcal{J}_1$, with $\mathcal{J}_{k}$ being defined as (see SI for details),

 \begin{align}
    \mathcal{J}_{k}=\begin{cases}
    \overline{J}^{(1)}_{k,(k-1)}~~~\text{ (Scenario I),}\\
    \\
    {J}^{(1)}_{k,(k-1)}~~~\text{ (Scenario II).}\\
    \end{cases}
\end{align}
In the SI, we numerically show that, not only the power in Scenario I is enhanced due to temperature fluctuations when the cold reservoir is finite ($\rho_1\neq 0$), but the efficiency of the heat engine is also affected depending on the ratio $V_1/V_2$.

\begin{figure}[b]
    \centering
        \includegraphics[width=1.0\linewidth]{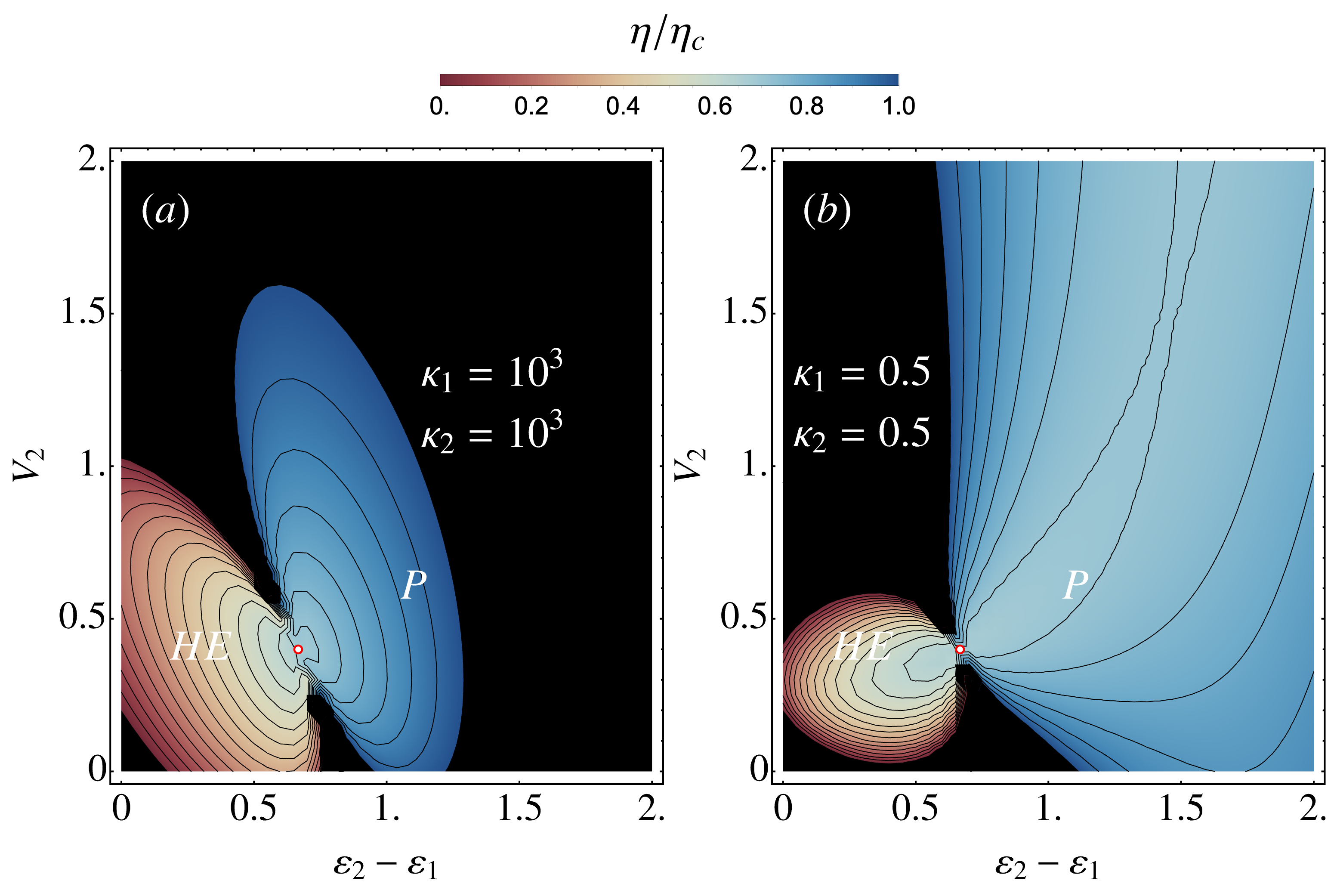}
    \caption{Efficiency heat maps for distinct values of $\kappa_1$ and $\kappa_2$ in Scenario II.
    Red circles mark the crossover from heat engine  to heat pump regimes through 
    the ideal efficiency, $\eta/\eta_c=1$. Parameters: $T_2=2T_1=1$ and $V_1=0.2$. }
    \label{fig_Heat_Maps}
\end{figure}

{In relation to Scenario II, we plot efficiency heat maps for large values of the thermal conductivities, $\kappa_1=\kappa_2=10^3$, for which the reservoirs are effectively of infinite size, in Fig.~\ref{fig_Heat_Maps}(a). We compare this with a case where both reservoirs are of finite-size, i.e., for $\kappa_1=\kappa_2=0.5$, plotted in Fig.~\ref{fig_Heat_Maps}(b). We see that in the latter case, the heat engine regime is shortened when compared to the infinite reservoir case. However, we also see that the heat pump engine regime is significantly enlarged in the finite-size reservoir case. This shows that, in the case of an engine composed of more than one interacting unit, finite reservoirs can greatly broaden the parameter regime wherein the machine operates as a heat pump.}
We note that the system operates at ideal efficiency $\eta=\eta_c$ when  $\varepsilon_1/\varepsilon_2=V_1/V_2=T_1/T_2=r^*$,
irrespective of the values of $\kappa_1$ and $\kappa_2$. In Fig.~\ref{fig_Heat_Maps}, this situation corresponds to the white cicles with red borders. 
 This is a direct consequence of Eq.~(\ref{ODETemp}): since $\mom{\dot{Q}_2},\langle {\cal P}\rangle\rightarrow 0$ in such case, the reservoir's finiteness is not relevant in the ideal regime.

Lastly, we investigate the effect of the reservoir's finiteness upon optimized quantities in Scenario II. 
It is known that the Carnot efficiency is impractical not only due to the difficulty of projecting an engine to operate ideally, but also because it implies operation at null power.
\begin{figure}
    \centering
        \includegraphics[width=0.49\linewidth]{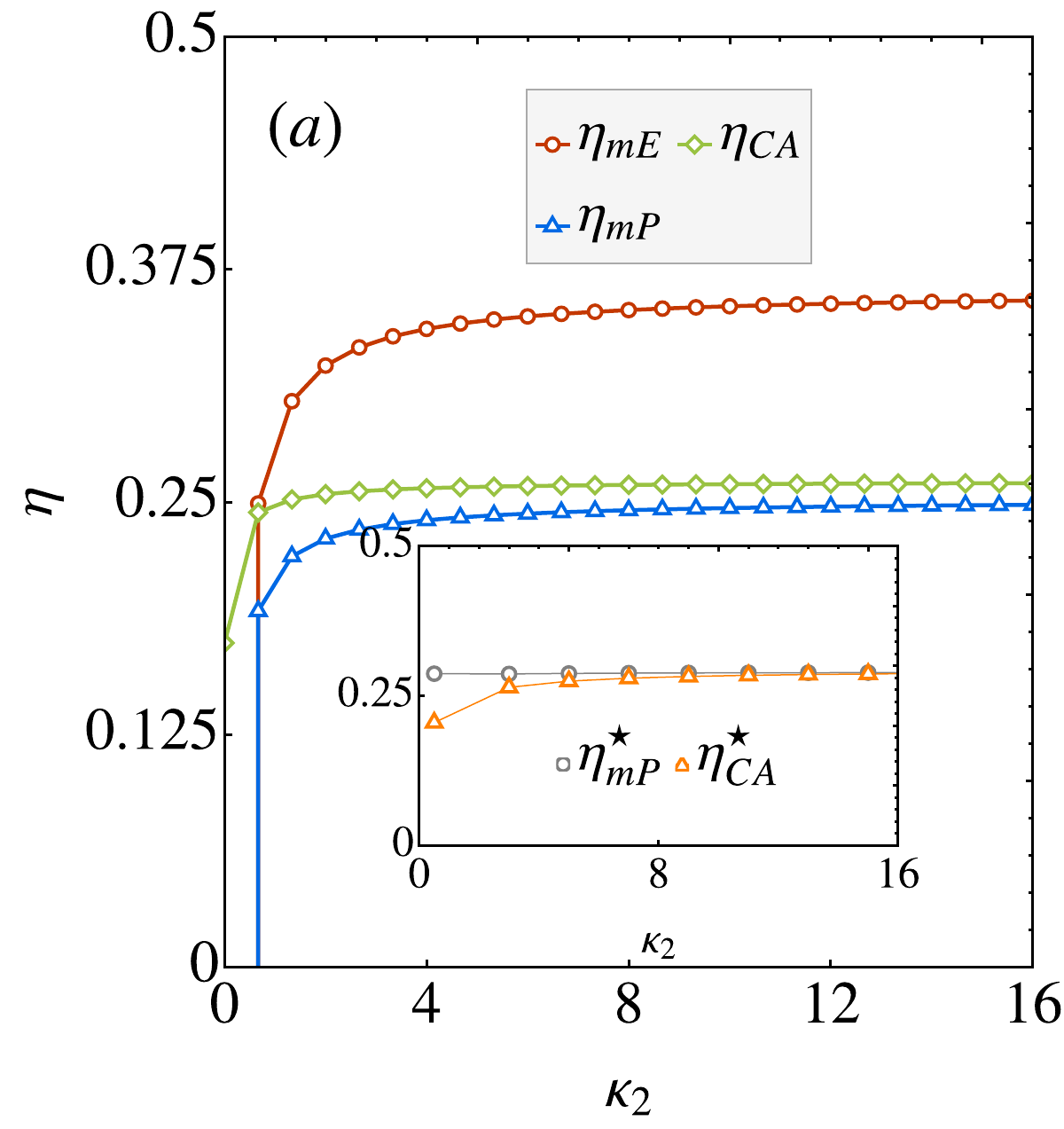}
    \includegraphics[width=0.49\linewidth]{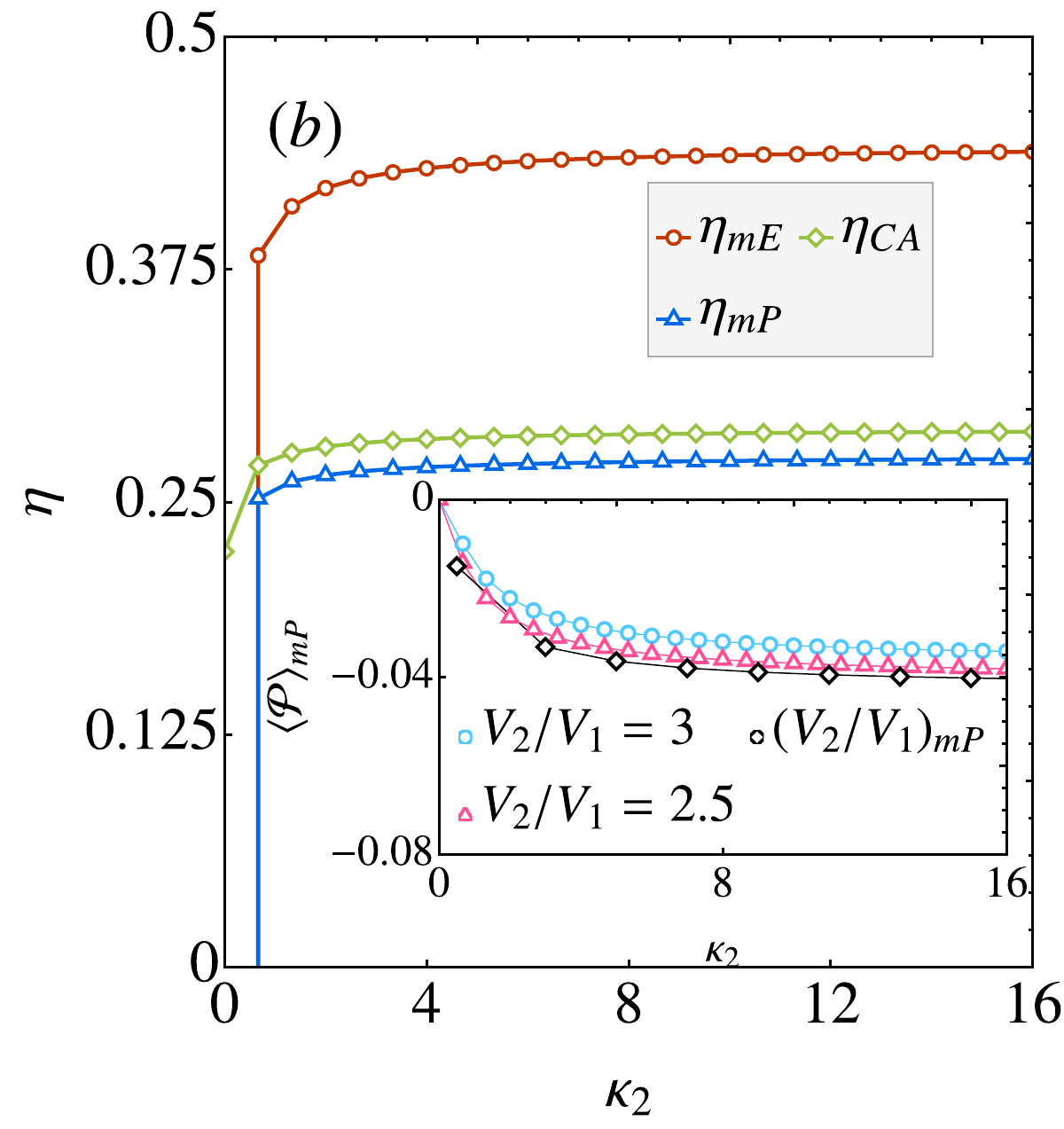}
    \caption{Optimized power and efficiencies as a function of $\kappa_2$ for $\kappa_1\rightarrow \infty$ in Scenario II.
    Panels (a) and (b) depict the maximum efficiency $\eta_{mE}$ and the efficiency at maximum power $\eta_{mP}$ for $V_2/V_1=3$ and $V_2/V_1=2.5$, respectively. {Insets: The depiction of $\eta^{\star}_{mP}$ and  $\mom{\mathcal{P}}_{mP}$ (right) versus $\kappa_2$, respectively.}}
    \label{fig_Max}
\end{figure}
For this reason, the Curzon-Ahlborn efficiency~\cite{Curzon1975, VandenBroeck2005}, defined as ${\eta}_{CA}\equiv 1-\sqrt{T_1/T_2}$, which describes the efficiency at maximum power, is very relevant. 

In Fig.~\ref{fig_Max}(a),
we plot the maximum efficiency, $\eta_{mE}$, and the efficiency at maximum power, $\eta_{mP}$, as a function of $\kappa_2$ for $\kappa_1\rightarrow\infty$, obtained by maximizing the power and efficiency over $\varepsilon_2-\varepsilon_1$ with fixed $V_2/V_1$.
We also plot the Curzon-Ahlborn efficiency, $\eta_{CA} = 1-\sqrt{T_1^*/T_2^*}$, where $T_1^*$ and $T_2^*$ are the stationary temperatures of the finite-size reservoirs for the values of $\varepsilon_2 - \varepsilon_1$ which maximize the power, for fixed $V_2/V_1$. 
In the inset, we also plot the efficiency maximized over both $\varepsilon_2-\varepsilon_1$ and $V_2/V_1$, which is given by $\eta_{mP}^\star$. In turn, the Curzon-Ahlborn efficiency $\eta_{CA}^\star$, is evaluated by using $T_1^*$ and $T_2^*$ for $\varepsilon_2-\varepsilon_1$ and $V_2/V_1$ that maximize the power. 
In Fig.~\ref{fig_Max}(b), we plot power at maximum efficiency $\langle {\cal P}\rangle_{mP}$ [pink and blue lines in the inset of panel~\ref{fig_Max}(b), with associated $\eta_{mP}$] and $\langle {\cal P^\star}\rangle_{mP}$ [black line in the inset of panel~\ref{fig_Max}(b), with associated $\eta^\star_{mP}$].

{In Figs.~\ref{fig_Max}(a)-(b), we see that, for small values of $\kappa_2$, the reservoir's finiteness
meaningfully influences all 
optimized quantities.
As $\kappa_2$ increases, all quantities in Fig.~\ref{fig_Max}(a)-(b) asymptotically approach their values for two infinite-size reservoirs.
Interestingly, by using the corrected temperatures in the finite-size case, for $\kappa_2$ small, we find that the Curzon-Ahlborn bound does not hold. While the Curzon-Ahlborn efficiency is already known to be non-universal and can be exceeded in certain heat engines with infinite-size reservoirs~\cite{Schmiedl2008, Apertet2017}, our work is the first to show that finite reservoirs also lead to violations of the Curzon-Ahlborn bound. As shown in the inset of Fig.~\ref{fig_Max}(a), the Curzon-Ahlborn efficiency is recovered in the limit of an infinite reservoir, establishing that the reservoir's finite size is the key ingredient for observing this effect here.}

\section{Conclusion}
We analyze the non-equilibrium thermodynamics of heat engines operating at finite-size reservoirs, which are in turn coupled to infinite-size reservoirs. 
We start from the more general description of 
heat engines composed of isothermal subprocesses operating at finite times and (instantaneous) adiabatic subprocesses, where the reservoir's finiteness is captured by fluctuating temperatures affecting the transition rates (Scenario I).
A complementary description is proposed to account for situations where the finite-size reservoirs are effectively simultaneously coupled to the system and to infinite-size reservoirs (Scenario II).
By utilizing the concept of entropic temperature, we derived an effective inverse temperature determining the average entropy production rate in Scenario I. 
In this case, depending on whether such inverse effective temperature leads to larger or lower temperature difference, one gets an improvement or deterioration of the extracted power. However, in Scenario II, we found that the temperature difference of the finite-size reservoirs always decrease in relation to the temperature difference of the associated infinite-size reservoirs, leading to a decrease in the average
extracted power.

Going beyond the two-state engine, which is characterized by only one independent flux, we analyzed a more complex three-state engine composed of two interacting units, whose performance also depends on the interplay between finite reservoirs and different
thermodynamic fluxes, showing that optimized power and efficiency are meaningfully affected by finite-size reservoirs in both scenarios.
Notably, this allowed us to show that the Curzon-Ahlborn bound does not always hold for finite-size reservoirs.
This work paves the way for a deeper understanding of heat engines operating at finite-size reservoirs. 
 {Although the influence of a finite reservoir is expected
to be less significant for sufficiently large systems,} it would be 
interesting to extend this analysis to more complex systems, such as those presenting collective behaviors and phase transitions coupled to finite-size reservoirs.

\section*{Acknowledgments}
We acknowledge Gabriel T. Landi for for insightful suggestions and inspiring discussions. The financial support from Brazilian agencies FAPESP and CNPq under grants 2023/17704-2, 2022/15453-0  and 2024/03763-0, respectively. M.T.M. is supported by a Royal Society University Research Fellowship. 
This project is co-funded by the European Union (Quantum Flagship project ASPECTS, Grant Agreement No.~101080167) and UK Research \& Innovation (UKRI). Views and opinions expressed are however those of the authors only and do not necessarily reflect those of the European Union, Research Executive Agency or UKRI. Neither the European Union nor UKRI can be held responsible for them.

\bibliography{refs}

\appendix

\onecolumngrid

\setcounter{equation}{0}
\setcounter{figure}{0}
\setcounter{table}{0}
\setcounter{page}{1}
\setcounter{section}{0}
\makeatletter
\renewcommand{\theequation}{A\arabic{equation}}
\renewcommand{\thefigure}{A\arabic{figure}}
\renewcommand{\citenumfont}[1]{A#1}



This Supplemental material  is structured as follows: Secs.  \ref{sc1} and \ref{sc2} present the main expressions and extra results for
Scenarios I and II, respectively.

\section{Thermodynamics of finite reservoirs: Scenario I} \label{sc1}

{As mentioned in the main text, in scenario I
the system is placed in contact with the cold (hot) finite reservoir during the time interval $0 < t \le \tau_1$ (stage 1), with the hot (cold) infinite reservoir when $\tau_1< t\leq \tau_1+\tau_2$ (stage 2) and  a periodic cycle is completed at $\tau = \tau_1 + \tau_2+\tau_3$.
Let $\nu$ the label for describing  the probability distribution at the $\nu$-th stage, its time evolution is governed by the  master equation:
\begin{equation}
    \centering
    \dot{p}^{(\nu)}_i(t)=\sum_{j\neq i}J^{(\nu)}_{ij},
    \label{me11}
\end{equation}
where $\nu=1(2)$ accounts to the cold (hot) thermal bath and $J^{(\nu)}_{ij}\equiv \omega^{(\nu)}_{ij}p^{(\nu)}_{j}(t)-\omega^{(\nu)}_{ji}p^{(\nu)}_{i}(t)$ is the corresponding probability current. As stated in the main text, one has
the following boundary conditions for the probability distribution:  $\vb{p}_i(0)=\vb{p}_i(\tau_1+\tau_2)$ and $\vb{p}_i(\tau_1)=\vb{p}_i(\tau_1)$, where $\vb{p}_\nu(t)\equiv(p^{(\nu)}_0(t), p^{(\nu)}_1(t), \dots, p^{(\nu)}_{N-1}(t))^{\rm T}$. The amount of power  in the first and second stages are  given by
${\mom{\mathcal{P}(t)}}=\sum_i(\varepsilon^{}_{2 i}-\varepsilon^{}_{1 i})\delta(t-\tau_1)p_i(t)$ and ${\mom{\mathcal{P}(t)}}=\sum_i(\varepsilon^{}_{1 i}-\varepsilon^{}_{2 i})\delta(t-\tau_2)p_i(t)$,  respectively, whose integration over a complete cycle together above boundary conditions read
${\mom{\cal P}}=-\sum_{i}\left(\varepsilon_{2i}-\varepsilon_{1i}\right)\left(p^{(1)}_i(\tau_1)-p^{(1)}_i(0)\right)/\tau$.} Additionally, integration of Eq.~(\ref{me11}) over
a complete cycle implies that  $\sum_{j\neq i}\overline{J}^{(1)}_{ij}=-\sum_{j\neq i}\overline{J}^{(2)}_{ij}$ for any $i$.


\subsection{General expressions for the two-state system}\label{apb}

\begin{figure}[htb!]
    \centering
    \includegraphics[width=0.6\linewidth]{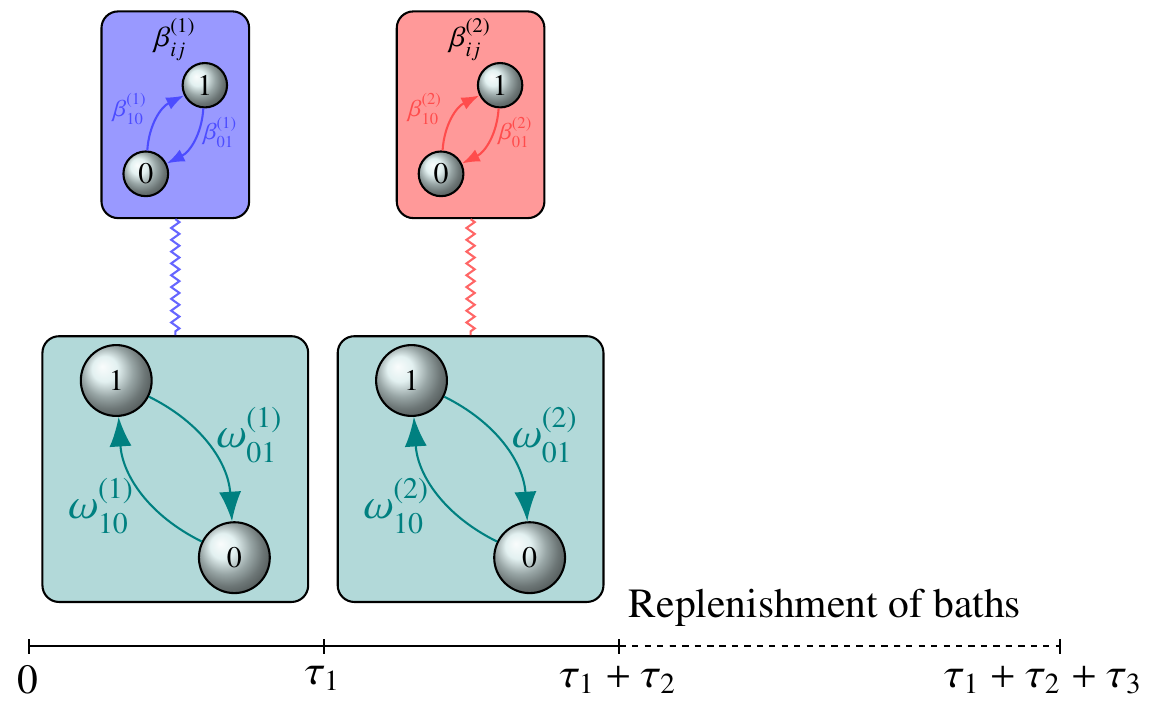}
    \caption{Schematics of dynamics and finite thermal reservoirs
    for the two-state system in scenario I.}
    \label{Schemes}
\end{figure}

{The two-state system has dynamics
 sketched in Fig.~\ref{Schemes}.}  
 The time evolution for the probability distribution is given by the following master equation
discussed in the main text. In Scenario I, transition rates will be re-expressed in the following form $\omega^{(\nu)}_{ij}\rightarrow   \Omega^{(\nu)}_{ij}$,
{\begin{equation}
    \Omega^{(\nu)}_{ij}=\omega^{(\nu)}_{(i,j|0)}p^{(\nu)}_0(\tau_1+\tau_2)+\omega^{(\nu)}_{(i,j|1)}p^{(\nu)}_1(\tau_1+\tau_2),
\end{equation}}
where $p^{(\nu)}_i(\tau_1+\tau_2)\equiv p_{i\nu}$ as discussed in the main text.
taking into account temperatures assuming different
values due to the fact the system can be empty or occupied
at $t=\tau_1+\tau_2$, where each  $\omega^{(\nu)}(i,j|k)$'s are given by
{\begin{align}
    \omega^{(\nu)}_{(1,0|0)}&=\Gamma\exp\left[-\frac{\beta_\nu}{2}\varepsilon_\nu\right],~~~~~\omega^{(\nu)}_{(1,0|1)}=\Gamma\exp\left[-\frac{\beta_\nu}{2(1+\rho_\nu)}\varepsilon_\nu\right],\\
    \omega^{(\nu)}_{(0,1|0)}&=\Gamma\exp\left[-\frac{\beta_\nu}{2(1-\rho_\nu)}\varepsilon_\nu\right],~~~~~\omega^{(\nu)}_{(0,1|1)}=\Gamma\exp\left[-\frac{\beta_\nu}{2}\varepsilon_\nu\right].
\end{align}}

As shown in the main text, the inverse temperatures  $\beta^{(\nu)}_{ij}$'s
are transition dependent and the time evolution of probability is then given by
{\begin{equation}
\label{mes}
    \centering
    \left(
\begin{array}{c}
 \dot{p}^{(\nu)}_0(t) \\
 \dot{p}^{(\nu)}_1(t) \\
\end{array}
\right)=\left(
\begin{array}{cc}
 -\Omega^{(\nu)}_{10} & \Omega^{(\nu)}_{01} \\
 \Omega^{(\nu)}_{10} & -\Omega^{(\nu)}_{01} \\
\end{array}
\right)\left(
\begin{array}{c}
 {p}^{(\nu)}_0(t) \\
 {p}^{(\nu)}_1(t) \\
\end{array}
\right),
\end{equation}}
{where  ${p}^{(\nu)}_0(t)=1-{p}^{(\nu)}_1(t)$ }.
Since we are interested in the time-dependent solution for applying the boundary conditions, the solution of the master equation is obtained using spectral decomposition of the transition matrix, such that the terms $p^{(\nu)}_0(t)$ and $p^{(\nu)}_1(t)$ are given by
{\begin{align}
    \centering
    p^{(\nu)}_0(t)&=\frac{\Omega^{(\nu)}_{10}}{\Omega^{(\nu)}_{10}+\Omega^{(\nu)}_{01}}+\frac{e^{-(\Omega^{(\nu)}_{10}+\Omega^{(\nu)}_{01})(t-\delta_{\nu,2}\tau_1)}[\Omega^{(\nu)}_{01}p^{(\nu)}_1(\delta_{\nu,2}\tau_1)-\Omega^{(\nu)}_{10}p^{(\nu)}_0(\delta_{\nu,2}\tau_1)]}{\Omega^{(\nu)}_{10}+\Omega^{(\nu)}_{01}},\\
    p^{(\nu)}_1(t)&=\frac{\Omega^{(\nu)}_{01}}{\Omega^{(\nu)}_{10}+\Omega^{(\nu)}_{01}}-\frac{e^{-(\Omega^{(\nu)}_{10}+\Omega^{(\nu)}_{01})(t-\delta_{\nu,2}\tau_1)}[\Omega^{(\nu)}_{01}p^{(\nu)}_1(\delta_{\nu,2}\tau_1)-\Omega^{(\nu)}_{10}p^{(\nu)}_0(\delta_{\nu,2}\tau_1)]}{\Omega^{(\nu)}_{10}+\Omega^{(\nu)}_{01}},
\end{align}}
where $\delta_{i,j}$ denotes the Kronecker delta and {using the boundary conditions } $p^{(1)}_0(0)=p^{(2)}_0(\tau_1+\tau_2)$ and $p^{(1)}_0(\tau_1)=p^{(2)}_0(\tau_1)$. {By inserting expressions
for transition rates, together above boundary conditions, expressions
for $ p^{(1)}_i(t)$ and  $ p^{(2)}_i(t)$ is obtained.} 
The integration of Eq.~(\ref{mes}) over each stage leads to the expressions

{
\begin{align}
    \overline{J}^{(1)}_{10}&=\frac{1}{\tau}\int_{0}^{\tau_1}dt~[\Omega^{(1)}_{10}p^{(1)}_0(t)-\Omega^{(1)}_{01}p^{(1)}_1(t)]=\frac{\tau_1(\Omega^{(1)}_{10}-\Omega^{(1)}_{01})}{\tau}-\frac{e^{-\tau_1(\Omega^{(1)}_{10}+\Omega^{(1)}_{01})}\left[1-e^{-{\tau_1(\Omega^{(1)}_{10}+\Omega^{(1)}_{01})}}\right][\Omega^{(1)}_{01}p^{(1)}_1(0)-\Omega^{(1)}_{10}p^{(1)}_0(0)]}{\tau\left(\Omega^{(1)}_{10}+\Omega^{(1)}_{01}\right)},\\
    \overline{J}^{(2)}_{10}&=\frac{1}{\tau}\int_{\tau_1}^{\tau_1+\tau_2}dt~[\Omega^{(2)}_{10}p^{(2)}_0(t)-\Omega^{(2)}_{01}p^{(2)}_1(t)]=\frac{\tau_2(\Omega^{(2)}_{10}-\Omega^{(2)}_{01})}{\tau}-\frac{e^{-\tau_2(\Omega^{(2)}_{10}+\Omega^{(2)}_{01})}\left[1-e^{-{\tau_2(\Omega^{(2)}_{10}+\Omega^{(2)}_{01})}}\right][\Omega^{(2)}_{01}p^{(2)}_1(\tau_1)-\Omega^{(2)}_{10}p^{(2)}_0(\tau_1)]}{\tau\left(\Omega^{(2)}_{10}+\Omega^{(2)}_{01}\right)},
\end{align}
}
where $\overline{J}^{(2)}_{10}=-\overline{J}^{(1)}_{10}$. Thermodynamic quantities in such case are promptly obtained and given by $ \mom{\dot{Q}_2}=\varepsilon_2\overline{J}^{(1)}_{10}$, $\langle {\cal P}\rangle=(\varepsilon_1-\varepsilon_2)\overline{J}^{(1)}_{10}$ and
$\langle{\dot \sigma}\rangle$ expressed in terms of transition rates:

{\begin{align}
    \mom{\dot{Q}_\nu}&=(-1)^{\nu+1}\varepsilon_\nu\left\{\frac{\tau_1(\Omega^{(1)}_{10}-\Omega^{(1)}_{01})}{\tau}-\frac{e^{-\tau_1(\Omega^{(1)}_{10}+\Omega^{(1)}_{01})}\left[1-e^{-{\tau_1(\Omega^{(1)}_{10}+\Omega^{(1)}_{01})}}\right][\Omega^{(1)}_{01}p^{(1)}_1(0)-\Omega^{(1)}_{10}p^{(1)}_0(0)]}{\tau\left(\Omega^{(1)}_{10}+\Omega^{(1)}_{01}\right)}\right\},\\
    \mom{\mathcal{P}}&=(\varepsilon_2-\varepsilon_1)\left\{\frac{\tau_1(\Omega^{(1)}_{10}-\Omega^{(1)}_{01})}{\tau}-\frac{e^{-\tau_1(\Omega^{(1)}_{10}+\Omega^{(1)}_{01})}\left[1-e^{-{\tau_1(\Omega^{(1)}_{10}+\Omega^{(1)}_{01})}}\right][\Omega^{(1)}_{01}p^{(1)}_1(0)-\Omega^{(1)}_{10}p^{(1)}_0(0)]}{\tau\left(\Omega^{(1)}_{10}+\Omega^{(1)}_{01}\right)}\right\}\\
    \mom{\dot{\sigma}}&=-\beta^{(1)}_{\rm eff}\mom{\dot{Q}_1}-\beta^{(2)}_{\rm eff}\mom{\dot{Q}_2},
    \end{align}}
    where $\beta^{(\nu)}_{\rm eff}$ is the entropic temperature discussed in the main text.


The engine regime is delimited by $\Delta\varepsilon=\varepsilon_2-\varepsilon_1=0$
and $(\Delta\varepsilon)_{\sigma}$, such latter limit being reservoir dependent.
Similar expressions
can be obtained for the finite reservoir composed of non-interacting units.
Since they are  more cumbersome, they will not be shown here.

\subsection{Beyond the two-state system: A
 minimal interacting heat engine}\label{tsm}

As stated in the main text, our second system is composed of three different states and also placed in contact with sequentially placed hot and a cold thermal bath whose fluctuating temperatures were stated in the main text. The states are composed by individual energies $0$, $\varepsilon_\nu+V_\nu$ and $2\varepsilon_\nu$, where $V_\nu$
is the  interacting term. It is worth mentioning that
only transitions $i\rightarrow i\pm 1$ are performed and hence
transitions $0\leftrightarrow 2$ are forbidden.  
In such case, the time evolution for the probability distribution 
of each state $i$ follows Eq.~(\ref{meMain}) given by
\begin{equation}
    \centering
    \left(
\begin{array}{c}
 \dot{p}^{(\nu)}_0(t) \\
 \dot{p}^{(\nu)}_1(t) \\
 \dot{p}^{(\nu)}_2(t)
\end{array}
\right)=\left(
\begin{array}{ccc}
 -\Omega^{(\nu)}_{10} & \Omega^{(\nu)}_{01} & 0 \\
 \Omega^{(\nu)}_{10} & -\Omega^{(\nu)}_{01}- \Omega^{(\nu)}_{21}& \Omega^{(\nu)}_{12} \\
 0 & \Omega^{(\nu)}_{21} & -\Omega^{(\nu)}_{12} \\
\end{array}
\right)\left(
\begin{array}{c}
 {p}^{(\nu)}_0(t) \\
 {p}^{(\nu)}_1(t) \\
 {p}^{(\nu)}_2(t)
\end{array}
\right),
\end{equation}
where{\begin{equation}
    \Omega^{(\nu)}_{ij}=\omega^{(\nu)}_{(i,j|0)}p^{(\nu)}_0(\tau_1+\tau_2)+\omega^{(\nu)}_{(i,j|1)}p^{(\nu)}_1(\tau_1+\tau_2)+\omega^{(\nu)}_{(i,j|2)}p^{(\nu)}_2(\tau_1+\tau_2),
\end{equation}} because the finite reservoir has energy $\mathcal{E}_\nu$ at $t=\tau$ and 
 ${p}^{(\nu)}_0(t)+{p}^{(\nu)}_1(t)+{p}^{(\nu)}_2(t)=1$ for any $t$. As a consequence, additional temperatures are also transition dependent and given by
\begin{align}
    \centering
    \beta^{(\nu)}_{01}&=\beta^{(\nu)}_{21}=\beta_\nu,\nonumber\\ \beta^{(\nu)}_{10}&=\beta_\nu\left(1+{\rho_\nu+\mathcal{V}_\nu}\right)^{-1},\\ 
    \beta^{(\nu)}_{12}&=\beta_\nu\left(1-{\rho_\nu+\mathcal{V}_\nu}\right)^{-1},\nonumber
\end{align}
provided $i=1$ at $t=\tau_1+\tau_2$, and 
\begin{align}
    \centering
    \beta^{(\nu)}_{12}&=\beta_\nu,\nonumber\\ \beta^{(\nu)}_{01}&=\beta^{(\nu)}_{21}=\beta_\nu\left(1+{\rho_\nu-\mathcal{V}_\nu}\right)^{-1},\\ 
    \beta^{(\nu)}_{10}&=\beta_\nu\left(1+{2\rho_\nu}\right)^{-1},\nonumber
\end{align}
provided $i=2$ at $t=\tau_1+\tau_2$. 

By proceeding as in the two-state model, expression 
for probability distribution is given by 
\begin{align}
    \centering
    p^{(\nu)}_{i}(t)&=\pi^{(\nu)}_i+\sum_{k=1}^{2}\sum_{j}e^{\lambda^{(\nu)}_k(t-\tau_{\nu-1})}(\bm{\phi}^{(\nu)}_k\otimes\bm{\psi}^{(\nu)}_k)_{ij}p_j^{(\nu)}(\tau_{\nu-1}),
\end{align}
where $\lambda_k^{(\nu)}$ is the second-largest eigenvalue with
corresponding eigenvectors  $\psi^{(\nu)}_k\cdot\phi^{(\nu')}_l=\delta_{k,l}\delta_{\nu,\nu'}$.
The expression for $\lambda^{(\nu)}_k$ is listed 
below
{\begin{equation}
    \centering
    \lambda^{(\nu)}_k=\Tr W_\nu\left\{\frac{1}{2}+(-1)^k\sqrt{1-4\left[\frac{\Omega^{(\nu)}_{01}\Omega^{(\nu)}_{12}+\Omega^{(\nu)}_{10}\Omega^{(\nu)}_{12}+\Omega^{(\nu)}_{10}\Omega^{(\nu)}_{21}}{\left(\Omega^{(\nu)}_{01}+\Omega^{(\nu)}_{10}+\Omega^{(\nu)}_{12}+\Omega^{(\nu)}_{21}\right)^2}\right]}\right\},
\end{equation}}
and, the steady-state probabilities are given by:
{\begin{equation}
    \pi^{(\nu)}_0=\frac{\Omega^{(\nu)}_{01}\Omega^{(\nu)}_{12}}{\Omega^{(\nu)}_{01}\Omega^{(\nu)}_{12}+\Omega^{(\nu)}_{10}\Omega^{(\nu)}_{12}+\Omega^{(\nu)}_{10}\Omega^{(\nu)}_{21}},~~~~\pi^{(\nu)}_1=\frac{\Omega^{(\nu)}_{10}\Omega^{(\nu)}_{12}}{\Omega^{(\nu)}_{01}\Omega^{(\nu)}_{12}+\Omega^{(\nu)}_{10}\Omega^{(\nu)}_{12}+\Omega^{(\nu)}_{10}\Omega^{(\nu)}_{21}},~~~~\pi^{(\nu)}_2=\frac{\Omega^{(\nu)}_{10}\Omega^{(\nu)}_{21}}{\Omega^{(\nu)}_{01}\Omega^{(\nu)}_{12}+\Omega^{(\nu)}_{10}\Omega^{(\nu)}_{12}+\Omega^{(\nu)}_{10}\Omega^{(\nu)}_{21}}.
\end{equation}}
From the above boundary conditions $p_i^{(\nu)}(0)=p_i^{(\nu)}(\tau_1+\tau_2)$ and $p_i^{(\nu)}(\tau_1)=p_i^{(\nu)}(\tau_1)$, each component $p^{(\nu)}_i(t)$
is obtained. As in the two-state system, by integrating the probability currents over a period, one obtains the independent fluxes
${\cal J}_{1}=\overline{J}^{(1)}_{10}$
and ${\cal J}_{2}=\overline{J}^{(1)}_{21}$:
{\begin{align}
    \centering
    \overline{J}^{(1)}_{10}&=\frac{1}{\tau}\int_{0}^{\tau_1}dt~[\Omega^{(1)}_{10}p^{(1)}_0(t)-\Omega^{(1)}_{01}p^{(1)}_1(t)],~~~~~\overline{J}^{(2)}_{10}=\frac{1}{\tau}\int_{\tau_1}^{\tau_1+\tau_2}dt~[\Omega^{(2)}_{10}p^{(2)}_0(t)-\Omega^{(2)}_{01}p^{(2)}_1(t)],\\
    \overline{J}^{(1)}_{21}&=\frac{1}{\tau}\int_{0}^{\tau_1}dt~[\Omega^{(1)}_{21}p^{(1)}_1(t)-\Omega^{(1)}_{12}p^{(1)}_2(t)],~~~~~\overline{J}^{(2)}_{21}=\frac{1}{\tau}\int_{\tau_1}^{\tau_1+\tau_2}dt~[\Omega^{(2)}_{21}p^{(2)}_1(t)-\Omega^{(2)}_{12}p^{(2)}_1(t)].
\end{align}}


{Moreover, the efficiency $\eta=-\mom{\mathcal{P}}/\mom{\dot{Q}_2}$ for this three-state model is given by:}
{\begin{align}
    \eta&=-\frac{(\varepsilon_1-\varepsilon_2)(\overline{J}^{(1)}_{10}-\overline{J}^{(1)}_{21})+(V_1-V_2)(\overline{J}^{(1)}_{10}+\overline{J}^{(1)}_{21})}{\varepsilon_2(\overline{J}^{(1)}_{10}-\overline{J}^{(1)}_{21})+V_2(\overline{J}^{(1)}_{10}+\overline{J}^{(1)}_{21})},
\end{align}}
alternatively expressed via ratio given by Eqs.~(\ref{A_Label}).

\section{Thermodynamics of finite reservoirs: Scenario II}\label{sc2}

\begin{figure}[htb!]
    \centering
    \includegraphics[width=0.6\linewidth]{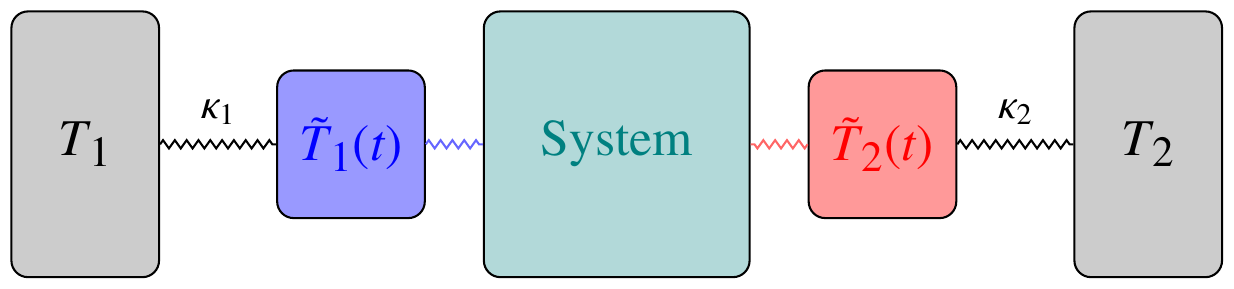}
    \caption{Schematics of dynamics and finite thermal reservoirs
    for a generic setup in scenario II. }
    \label{SchemeII}
\end{figure}

{{In scenario II, the finiteness of
reservoirs is studied via a simultaneous  contact
with an infinite reservoir. Among the distinct approaches, it is introduced through a } 
 heat flux proportional to
the difference of temperatures and thermal conductance $\kappa_\nu$, as sketched
in Fig.~\ref{SchemeII}.}

From the first law of Thermodynamics, the time variation of energy
of finite reservoir is given by  the exchanged heat between the system and the reservoir,  $\mom{\dot{Q}_\nu(t)}$,  and its coupling
with the infinite bath described before.  For constant heat capacity,
one has the following expression for the time evolution of $\tilde{T}_\nu$
\begin{equation}
    \centering
    C_\nu\dv{\tilde{T}_\nu(t)}{t}=-[\mom{\dot{Q}_\nu(t)}+\kappa_\nu(\tilde{T}_\nu(t)-T_\nu)],
    \label{resf}
\end{equation}
where $C_\nu$, $\tilde{T}_\nu(t)$ and $T_\nu$ are the heat capacity, temperature and the temperature of the infinite reservoir coupled to the $\nu$-th finite bath, respectively, and $\langle \dot{{Q}}_\nu\rangle = \sum_{(ij)}\Delta \mathcal{E}^{(\nu)}_{ij}J_{ij}^{(\nu)}(t)$, where
the time evolution of probabilities is governed by the master equation
\begin{equation}
    \centering
    \dot{p}_i(t)=\sum_{\nu}\sum_{j\neq i}[\omega_{ij}^{(\nu)}(t)p_j(t)-\omega_{ji}^{(\nu)}(t)p_i(t)]=\sum_{\nu}\sum_{j\neq i}J_{ij}^{(\nu)}(t).
    \label{ME}
\end{equation}
{ We are interested in the steady-state  regime in which
 $p_k(t)\rightarrow\pi_k$, $\tilde{T}_\nu(t)\rightarrow \tilde{T}_\nu^*$, obtained by solving Eqs.(\ref{resf})-(\ref{ME}).}
Having them, all thermodynamic quantities are promptly obtained
as described below
\begin{eqnarray}
\label{work}
 \langle \dot{\sigma} \rangle= \sum_{\nu}\sum_{(i<j)}J_{ij}^{(\nu)}\ln\frac{\omega^{(\nu)}_{ij}}{\omega^{(\nu)}_{ji}}=-\sum_{\nu}\frac{\langle \dot{{Q}}_\nu\rangle}{\tilde{T}_\nu^*},
\end{eqnarray}
where $\langle {\dot \sigma}\rangle\ge 0$ and the mean power  $\langle {\cal P}\rangle $ is given by
$\langle {\cal P}\rangle = -\sum_{(ij)}(\Delta \mathcal{E}^{(1)}_{ij}J_{ ij}^{(1)}+\Delta \mathcal{E}^{(2)}_{ij}J_{ij}^{(2)})$, whose relation with $\langle \dot{{Q}}_\nu\rangle$'s 
obeys  the first  law of Thermodynamics $\langle {\cal P} \rangle+\langle \dot{{Q}}_1\rangle+\langle\dot{{ Q}}_2\rangle=0$.

\subsection{General expressions for the two-state system}
\begin{figure}
    \centering
    \includegraphics[scale=0.25]{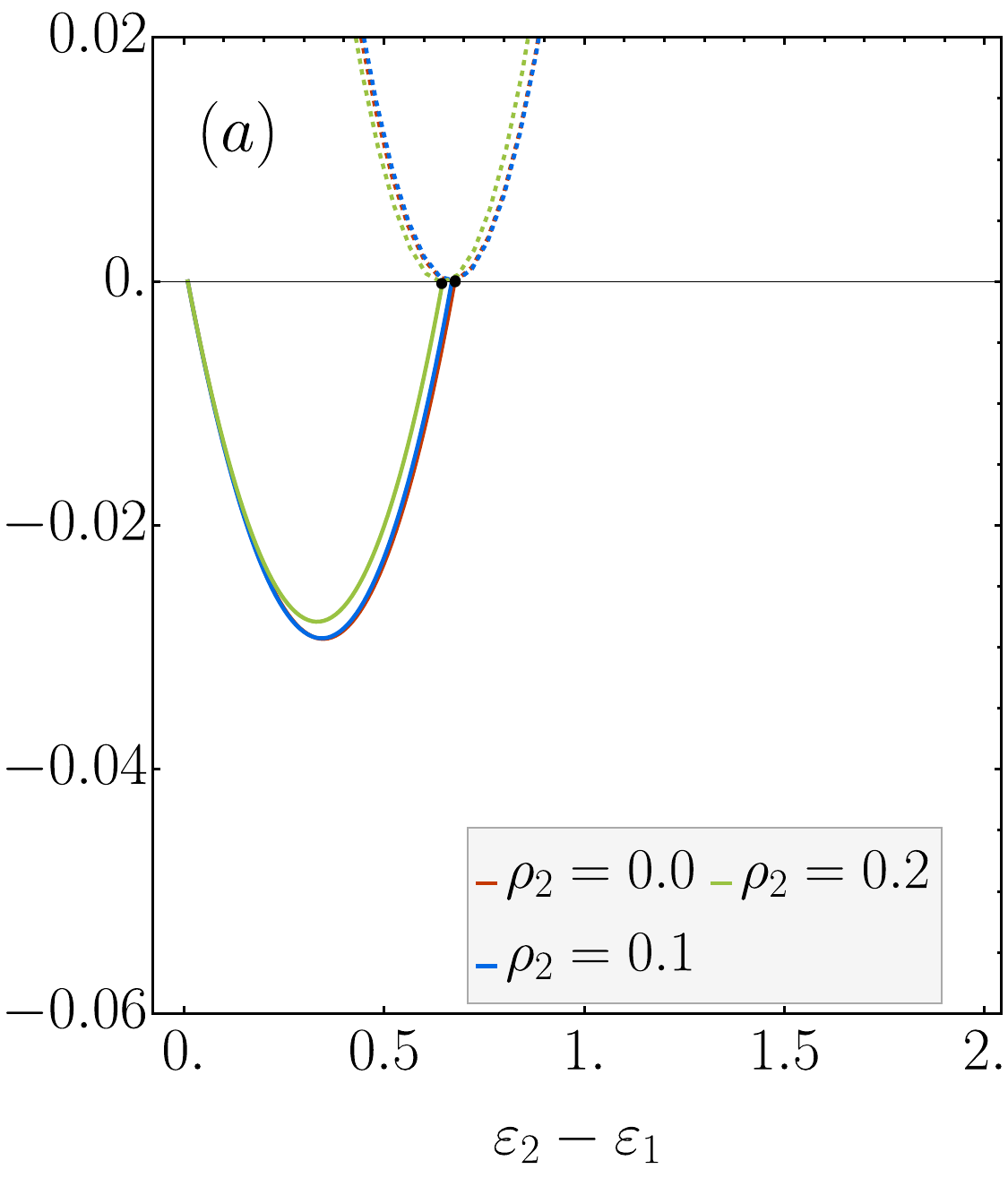}
    \includegraphics[scale=0.25]{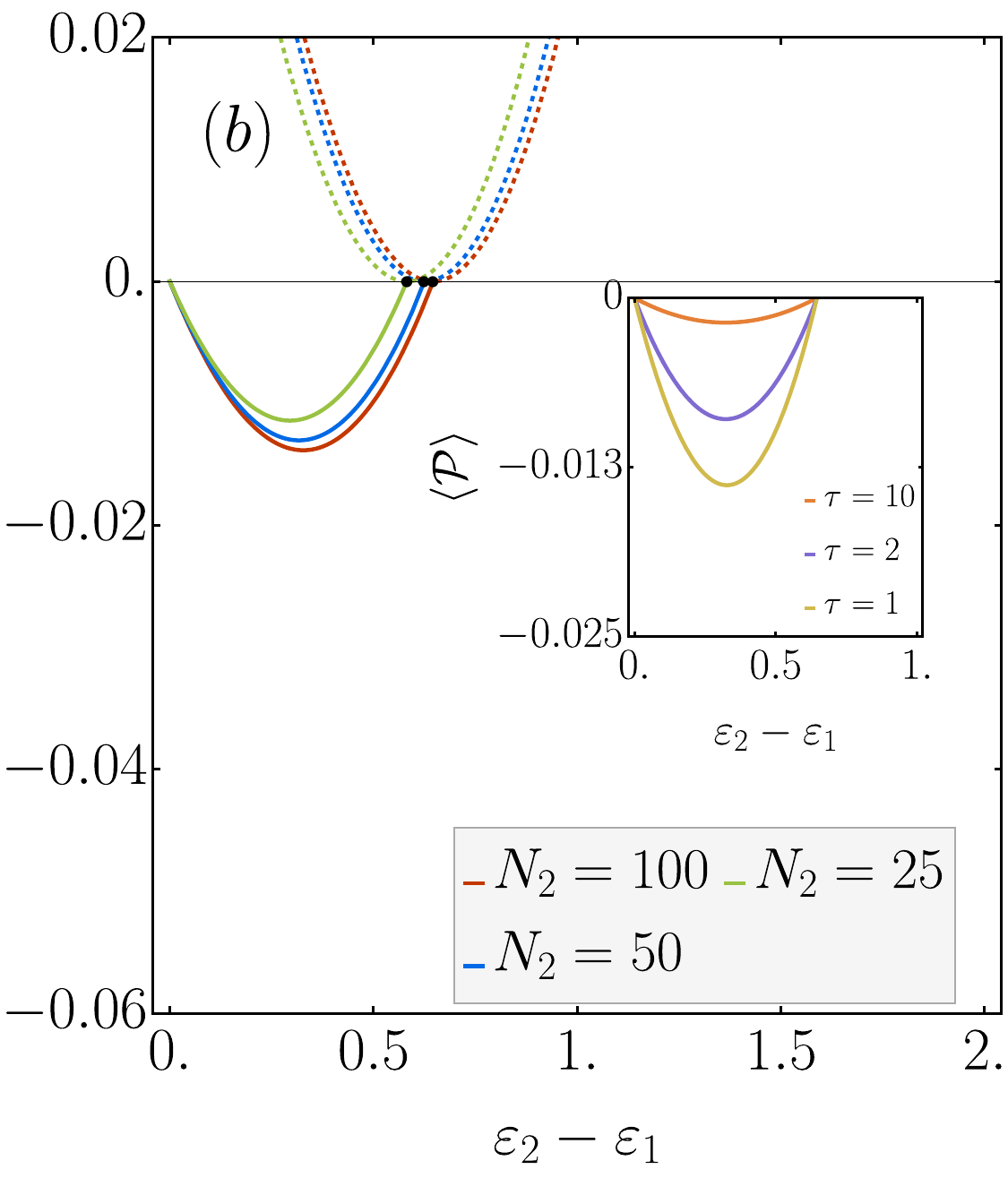}
    \includegraphics[scale=0.25]{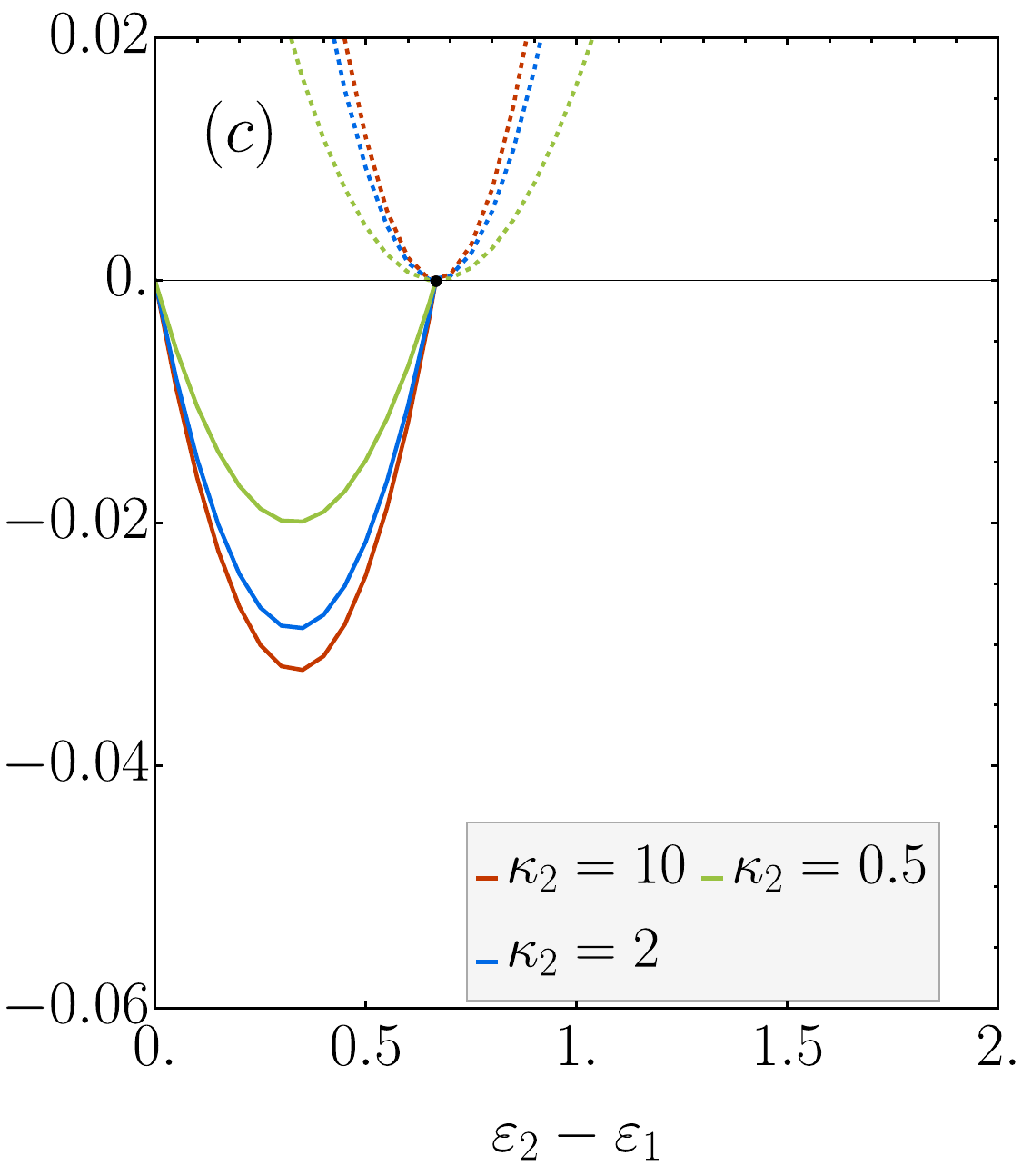}
    \caption{The same as in Fig.\ref{fig_General}, but for the finite hot reservoir.
}
    \label{fig_General_Hot}
\end{figure}

Starting with the former two-state model placed, the dynamics
is characterized by the following equations
$ dp_0(t)/dt=J^{(1)}_{01}(t)+J^{(2)}_{01}(t)$ and
\begin{align}\label{rsf1}
    C_1\dv{\tilde{T}_1(t)}{t}&=\varepsilon_1J^{(1)}_{01}(t)-\kappa_1(\tilde{T}_1(t)-T_1),\\
     C_2\dv{\tilde{T}_2(t)}{t}&=\varepsilon_2J^{(2)}_{01}(t)-\kappa_2(\tilde{T}_2(t)-T_2),\nonumber
\end{align}
where $\omega^{(\nu)}_{10}(t)=\Gamma e^{-\varepsilon_\nu/2\tilde{T}_\nu(t)}$, $\omega^{(\nu)}_{01}(t)=\Gamma e^{\varepsilon_\nu/2\tilde{T}_\nu(t)}$ and $J^{(\nu)}_{ij}(t)=\omega^{(\nu)}_{ij}p_j(t)-\omega^{(\nu)}_{ji}p_i(t)$. 
Expressions for  steady probabilities  $\{\pi_i\}$ are obtained via spanning-tree method \cite{Schnakenberg}  and given by
{\begin{align}
    \pi_0=\frac{\omega^{(1)}_{01}+\omega^{(2)}_{01}}{\omega^{(1)}_{01}+\omega^{(2)}_{01}+\omega^{(1)}_{10}+\omega^{(2)}_{10}},&&\pi_1=1-\pi_0=\frac{\omega^{(1)}_{10}+\omega^{(2)}_{10}}{\omega^{(1)}_{01}+\omega^{(2)}_{01}+\omega^{(1)}_{10}+\omega^{(2)}_{10}}.
\end{align}
By inserting them into the right side of Eq.~(\ref{rsf1}),
  steady temperatures $\tilde{T}^{*}_\nu$'s are obtained by solving
  the following system of non-linear equations
\begin{align}
    \varepsilon_1\tanh\left(\frac{\varepsilon_2}{2\tilde{T}_2^*}-\frac{\varepsilon_1}{2\tilde{T}_1^*}\right)=\kappa_1(T_1-\tilde{T}_1^*){\cosh\left(\frac{\varepsilon_2}{2\tilde{T}_2^*}+\frac{\varepsilon_1}{2\tilde{T}_1^*}\right)},&&    -\varepsilon_2\tanh\left(\frac{\varepsilon_2}{2\tilde{T}_2^*}-\frac{\varepsilon_1}{2\tilde{T}_1^*}\right)=\kappa_2(T_2-\tilde{T}_2^*){\cosh\left(\frac{\varepsilon_2}{2\tilde{T}_2^*}+\frac{\varepsilon_1}{2\tilde{T}_1^*}\right)}.
    \label{nonLinearSol}
\end{align}}
{Once again, thermodynamic quantities are promptly obtained via expressions $ \mom{\dot{Q}_2}=\varepsilon_2J^{(1)}_{10}$, $\langle {\cal P}\rangle=(\varepsilon_1-\varepsilon_2)J^{(1)}_{10}$ and
$\mom{\dot{\sigma}}=\langle {\cal P}\rangle+(1-{\tilde T^*_1}/{\tilde T^*_2})\langle {\dot Q}_2\rangle$, whose corresponding temperatures are given via the numerical solution of Eq.\eqref{nonLinearSol}. They become simpler for
strong $\kappa_\nu$'s, in which each  temperature is approximately expressed as $\tilde{T}^*_\nu=T_\nu+\delta T_\nu$, where $|\delta T_\nu|\ll 1$ and can be viewed as a linear correction with respect to the infinite case. By inserting into} Eq.~\eqref{nonLinearSol}, they are given by
\begin{align}
    \delta T_1=\frac{\varepsilon _1 \kappa _2 T_1^2 T_2^2 \sinh \left(2 \Omega _1\right)}{\kappa _1 T_1^2 \left(\varepsilon _2^2+2 \kappa _2 T_2^2 \cosh ^2\left(\Omega _1\right) \cosh
   \left(\Omega _2\right)\right)+\varepsilon _1^2 \kappa _2 T_2^2},&&
   \delta T_2=-\frac{\varepsilon _2 \kappa _1 T_1^2 T_2^2 \sinh \left(2 \Omega _1\right)}{\kappa _1 T_1^2 \left(\varepsilon _2^2+2 \kappa _2 T_2^2 \cosh ^2\left(\Omega _1\right) \cosh
   \left(\Omega _2\right)\right)+\varepsilon _1^2 \kappa _2 T_2^2},
\end{align}
{where $\Omega_k=(T_2\varepsilon_1+(-1)^kT_1\varepsilon_2)/4T_1T_2$. Since this approach is equivalent of an infinite model with $\beta_\nu\rightarrow1/\tilde{T}^*_\nu$, considering the limit of large values of $\kappa_\nu$, we extract the power $\mom{\mathcal{P}}$, with linear corrections, that is given by:}
{\begin{align}
    \mom{\mathcal{P}}&\approx\mom{\mathcal{P}}_\infty+\frac{\left(\varepsilon _2-\varepsilon _1\right)}{4}\left(\frac{\varepsilon _1^2 \cosh \left(\frac{\varepsilon _2}{2 T_2}\right)}{\kappa _1 T_1^2}+\frac{\varepsilon _2^2 \cosh \left(\frac{\varepsilon _1}{2 T_1}\right)}{\kappa _2 T_2^2}\right)  \tanh \left[\frac{1}{2} \left(\frac{\varepsilon _1}{T_1}-\frac{\varepsilon _2}{T_2}\right)\right]\text{sech}^2\left[\frac{1}{4}
   \left(\frac{\varepsilon _1}{T_1}+\frac{\varepsilon _2}{T_2}\right)\right] \text{sech}\left[\frac{1}{2} \left(\frac{\varepsilon _1}{T_1}+\frac{\varepsilon _2}{T_2}\right)\right]
\end{align}}
{The depiction of the two-state case for both approaches is shown in Fig.~\ref{fig_General_Hot} together Fig.\ref{fig_General}.}

\subsection{General expressions for the minimal interacting engine }
In a similar fashion, the dynamics of the three-state system is described via the system of  equations:
\begin{equation}
\begin{split}
    \dv{p_0(t)}{t}&=J^{(1)}_{01}(t)+J^{(2)}_{01}(t)\\
    \dv{p_2(t)}{t}&=J^{(1)}_{21}(t)+J^{(2)}_{21}(t)\\
    \end{split}
\end{equation}
and
\begin{equation}\label{rsf}
\begin{split}
    C_1\dv{\tilde{T}_1(t)}{t}&=(V_1+\varepsilon_1)J^{(1)}_{01}(t)+(V_1-\varepsilon_1)J^{(1)}_{21}(t)-\kappa_1(\tilde{T}_1(t)-T_1)\\
   C_2 \dv{\tilde{T}_2(t)}{t}&=-(V_2+\varepsilon_2)J^{(1)}_{01}(t)-(V_2-\varepsilon_2)J^{(1)}_{21}(t)-\kappa_2(\tilde{T}_2(t)-T_2)
\end{split}
\end{equation}
where $\omega^{(\nu)}_{10}=2\Gamma e^{-(\varepsilon_\nu+V_\nu)/(2\Tilde{T}_\nu(t))}$, $\omega^{(\nu)}_{21}=\Gamma e^{-(\varepsilon_\nu-V_\nu)/(2\Tilde{T}_\nu(t))}$, $\omega^{(\nu)}_{01}=\Gamma e^{(\varepsilon_\nu+V_\nu)/(2\Tilde{T}_\nu(t))}$ and $\omega^{(\nu)}_{12}=2\Gamma e^{(\varepsilon_\nu+V_\nu)/(2\Tilde{T}_\nu(t))}$. 
As previously, expressions for  steady probabilities  $\{\pi_i\}$ are obtained via spanning-tree method and given by
\begin{align}
    \pi_0&=\frac{(\omega^{(1)}_{12}+\omega^{(2)}_{12})(\omega^{(1)}_{01}+\omega^{(2)}_{01})}{(\omega^{(1)}_{12}+\omega^{(2)}_{12})(\omega^{(1)}_{01}+\omega^{(2)}_{01})+(\omega^{(1)}_{12}+\omega^{(2)}_{12})(\omega^{(1)}_{10}+\omega^{(2)}_{10})+(\omega^{(1)}_{21}+\omega^{(2)}_{21})(\omega^{(1)}_{01}+\omega^{(2)}_{21})},\\
    \pi_1&=\frac{(\omega^{(1)}_{12}+\omega^{(2)}_{12})(\omega^{(1)}_{10}+\omega^{(2)}_{10})}{(\omega^{(1)}_{12}+\omega^{(2)}_{12})(\omega^{(1)}_{01}+\omega^{(2)}_{01})+(\omega^{(1)}_{12}+\omega^{(2)}_{12})(\omega^{(1)}_{10}+\omega^{(2)}_{10})+(\omega^{(1)}_{21}+\omega^{(2)}_{21})(\omega^{(1)}_{01}+\omega^{(2)}_{21})},\\
    \pi_2&=\frac{(\omega^{(1)}_{21}+\omega^{(2)}_{21})(\omega^{(1)}_{01}+\omega^{(2)}_{21})}{(\omega^{(1)}_{12}+\omega^{(2)}_{12})(\omega^{(1)}_{01}+\omega^{(2)}_{01})+(\omega^{(1)}_{12}+\omega^{(2)}_{12})(\omega^{(1)}_{10}+\omega^{(2)}_{10})+(\omega^{(1)}_{21}+\omega^{(2)}_{21})(\omega^{(1)}_{01}+\omega^{(2)}_{21})}.
\end{align}
By inserting them into the right side of Eq.~(\ref{rsf}),
  steady temperatures $\tilde{T}^{*}_\nu$'s are obtained by solving
  the following system of non-linear equations
{\begin{align}
    \frac{4 e^{\frac{1}{2} \left(\frac{\varepsilon _1}{\tilde{T}^*_1}+\frac{\varepsilon _2}{\tilde{T}^*_2}\right)} \left[e^{\frac{V_1}{2 \tilde{T}^*_1}} \left(\varepsilon _1 \sinh \left(\frac{\varepsilon
   _1}{2 \tilde{T}^*_1}\right)+V_1 \cosh \left(\frac{\varepsilon _1}{2 \tilde{T}^*_1}\right)\right)-e^{\frac{V_2}{2 \tilde{T}^*_2}} \left(\varepsilon _1 \sinh \left(\frac{\varepsilon _2}{2
   \tilde{T}^*_2}\right)+V_1 \cosh \left(\frac{\varepsilon _2}{2 \tilde{T}^*_2}\right)\right)\right]}{2 e^{\frac{1}{2} \left(\frac{\varepsilon _1}{\tilde{T}^*_1}+\frac{\varepsilon
   _2}{\tilde{T}^*_2}\right)}+e^{\frac{1}{2} \left(\frac{2 \varepsilon _1+V_1}{\tilde{T}^*_1}+\frac{2 \varepsilon _2+V_2}{\tilde{T}^*_2}\right)}+e^{\frac{1}{2}
   \left(\frac{V_1}{\tilde{T}^*_1}+\frac{V_2}{\tilde{T}^*_2}\right)}}&=\kappa_1(\tilde{T}_1^*-T_1)\label{T1Three}\\
    -\frac{4 e^{\frac{1}{2} \left(\frac{\varepsilon _1}{\tilde{T}^*_1}+\frac{\varepsilon _2}{\tilde{T}^*_2}\right)} \left[e^{\frac{V_1}{2 \tilde{T}^*_1}} \left(\varepsilon _2 \sinh \left(\frac{\varepsilon
   _1}{2 \tilde{T}^*_1}\right)+V_2 \cosh \left(\frac{\varepsilon _1}{2 \tilde{T}^*_1}\right)\right)-e^{\frac{V_2}{2 \tilde{T}^*_2}} \left(\varepsilon _2 \sinh \left(\frac{\varepsilon _2}{2
   \tilde{T}^*_2}\right)+V_2 \cosh \left(\frac{\varepsilon _2}{2 \tilde{T}^*_2}\right)\right)\right]}{2 e^{\frac{1}{2} \left(\frac{\varepsilon _1}{\tilde{T}^*_1}+\frac{\varepsilon
   _2}{\tilde{T}^*_2}\right)}+e^{\frac{1}{2} \left(\frac{2 \varepsilon _1+V_1}{\tilde{T}^*_1}+\frac{2 \varepsilon _2+V_2}{\tilde{T}^*_2}\right)}+e^{\frac{1}{2}
   \left(\frac{V_1}{\tilde{T}^*_1}+\frac{V_2}{\tilde{T}^*_2}\right)}}&=\kappa_2(\tilde{T}_2^*-T_2),\label{T2Three}
\end{align}}
respectively.

\begin{figure}
    \centering
    \includegraphics[scale=0.35]{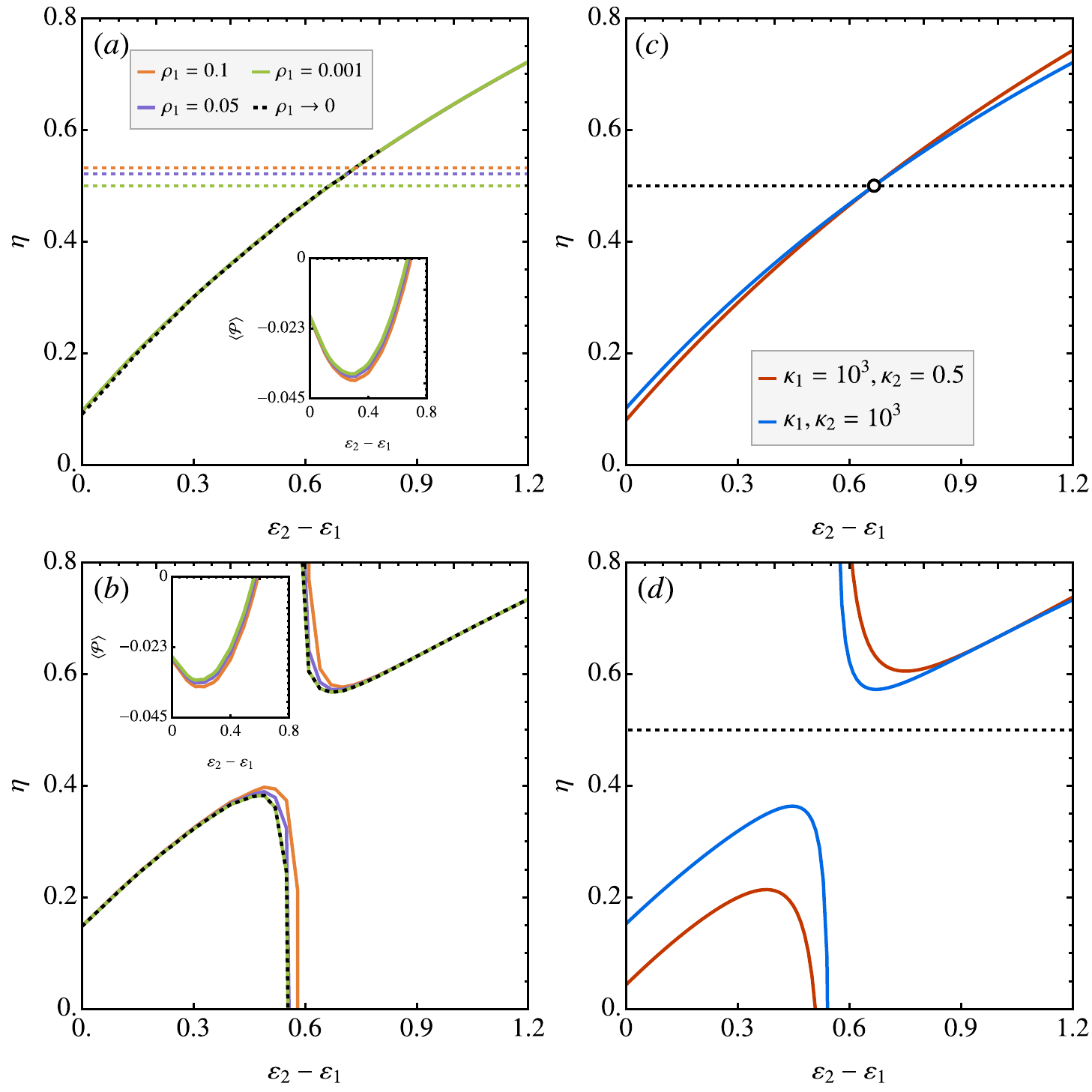}
    \caption{Panels $(a)-(b)$  and $(c)-(d)$ depict the efficiency $\eta$ versus $\varepsilon_2-\varepsilon_1$ 
    for scenarios I and II, respectively, for $V_1/V_2=1/2$ (top) and $1/3$ (bottom). Horizontal dashed lines
    denote corresponding Carnot efficiencies.
}
    \label{figa2}
\end{figure}

\begin{figure}[htb!]
    \centering
    \includegraphics[width=0.6\linewidth]{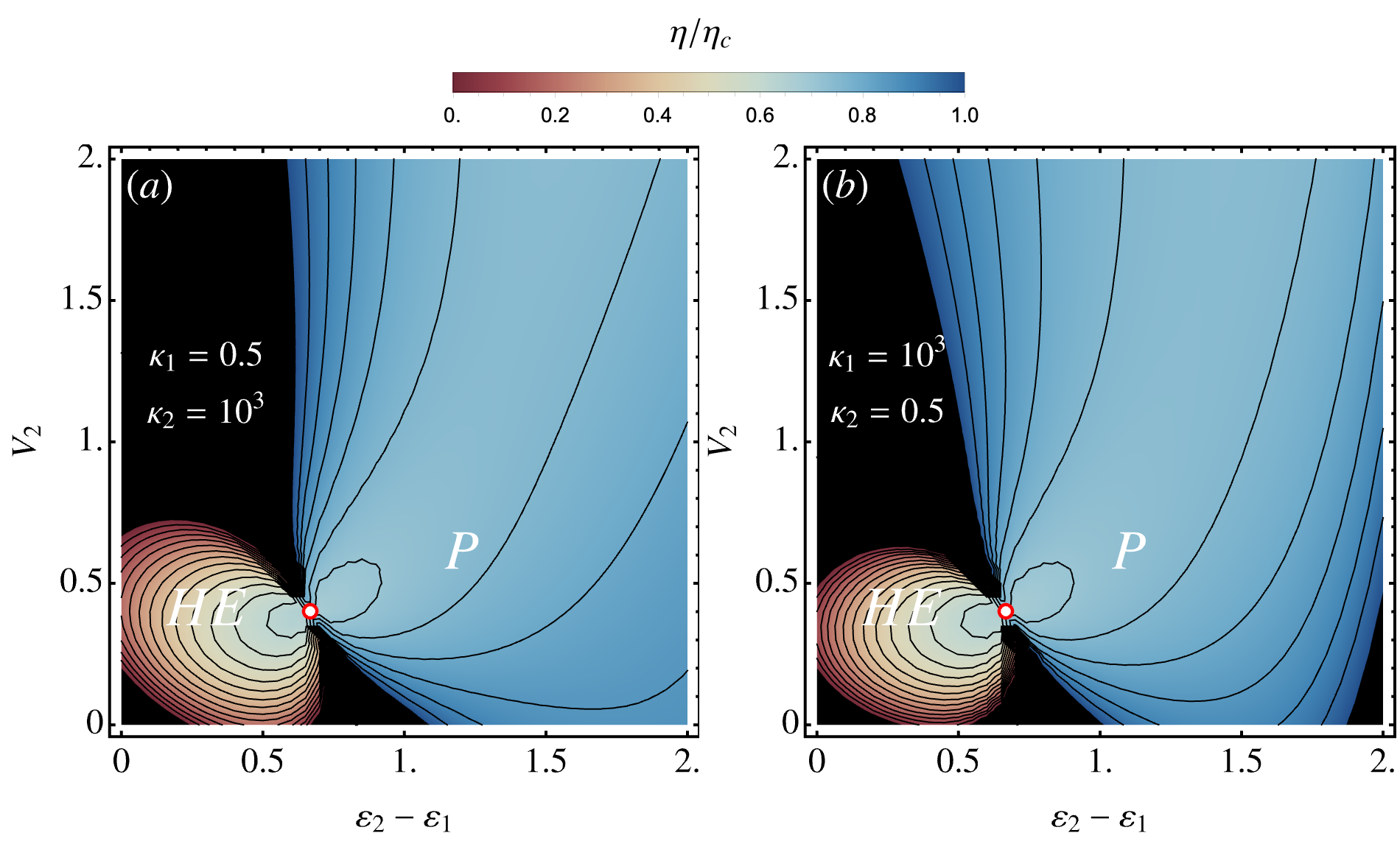}
    \caption{The same as Fig.\ref{fig_Heat_Maps} but for
    intermediate sets of  $\kappa_\nu$'s.}
    \label{fig_Heat_Maps2}
\end{figure}

\begin{figure}
    \centering
    \includegraphics[width=0.5\linewidth]{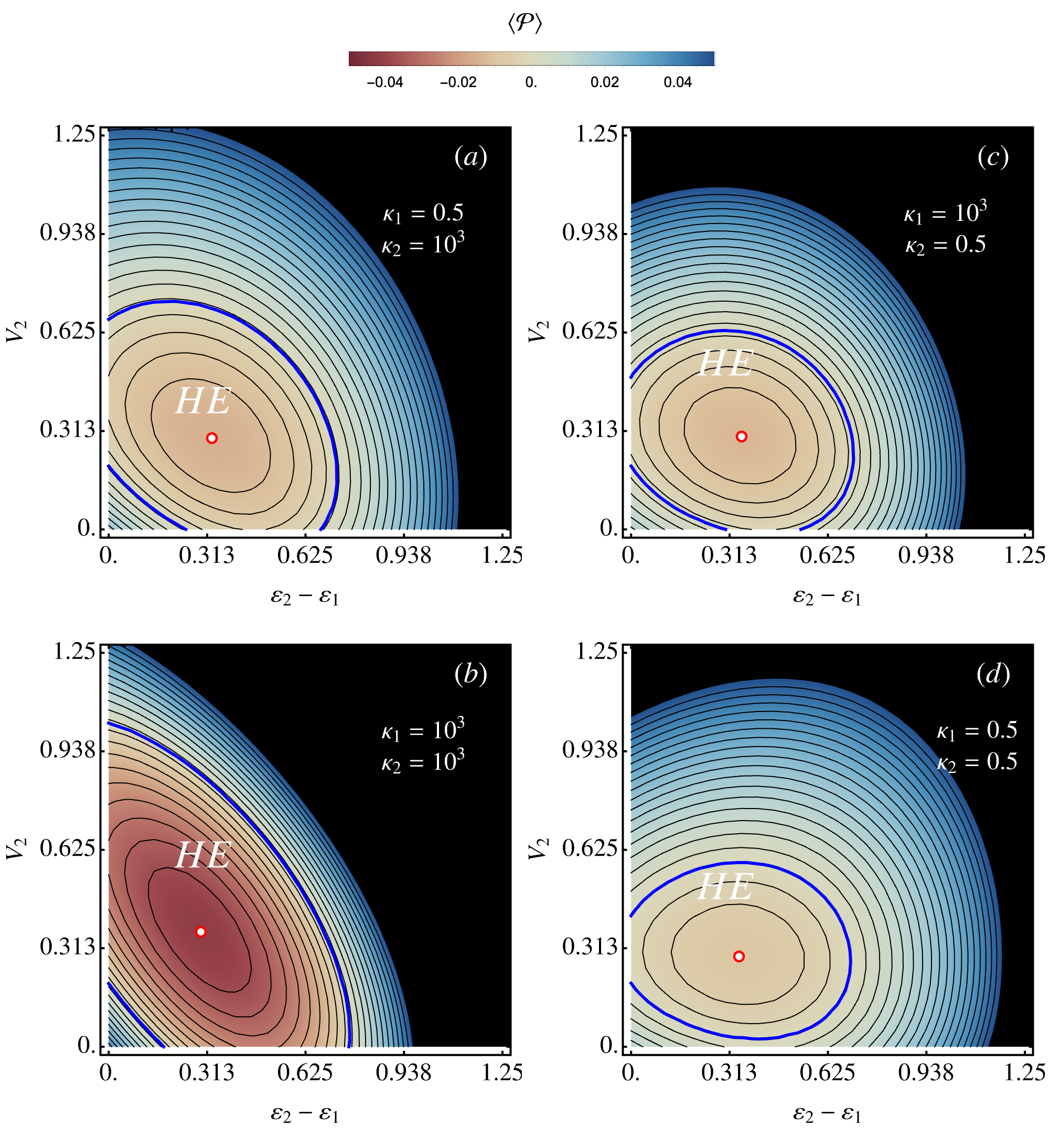}
    \caption{For the same parameters of Figs.\ref{fig_Heat_Maps} and \ref{fig_Heat_Maps2}, the corresponding power-output heat maps. Blue contours separates the heat-engine and the heat pump regimes.
    }
    \label{fig_Heat_Maps_Power}
\end{figure}

{Likewise to Fig.\ref{fig_General},
Fig.~\ref{figa2} draws a comparison between Scenarios I and II, but
for the efficiency $\eta$. As can be seen, both scenarios
provide similar findings about the role of reservoir finiteness.
Finally, Figs.~\ref{fig_Heat_Maps2} and \ref{fig_Heat_Maps_Power} depict the remaining efficiencies and corresponding power-output heat maps of Fig.~\ref{fig_Heat_Maps}, respectively.}

\end{document}